\documentclass[11pt,a4paper]{article}
\usepackage{jheppub}
\usepackage{natbib}
\usepackage{graphicx}
\usepackage[dvipsnames]{xcolor}

\usepackage{mathtools}
\usepackage{hyperref}
\usepackage{float}
\usepackage{comment}
\usepackage[title]{appendix}

\newcommand{\beq}{\begin{equation} }
\newcommand{\eeq}{\end{equation}} 

\definecolor{Blu}{rgb}{0.,0.,1.}

\title{Searching for  relativistic axions in the sky}

\author[a,b]{Arpan Kar,} \author[c]{Tanmoy Kumar,} \author[c]{Sourov Roy,} \author[d]{and Jure Zupan}

\affiliation[a]{Center for Quantum Spacetime, Sogang University, Seoul 121-742, South Korea}
\affiliation[b]{Department of Physics, Sogang University, Seoul 121-742, South Korea}
\affiliation[c]{School of Physical Sciences, Indian Association for the Cultivation of Science,\\ 2A \& 2B Raja S.C. Mullick Road, Jadavpur, Kolkata 700032, India}

\affiliation[d]{Department of Physics, University of Cincinnati, Cincinnati, Ohio 45221,USA
}  
  
\emailAdd{arpankarphys@gmail.com}
\emailAdd{kumartanmoy1998@gmail.com} 
\emailAdd{tpsr@iacs.res.in}
\emailAdd{zupanje@ucmail.uc.edu}

\abstract{Relativistic axions produced in decays of 
${\mathcal O}(10^{-7}-10^{-2}$ $\text{eV})$ dark matter (DM) partially convert to photons after 
traversing the galactic magnetic field, giving rise to a signal observable by the Square Kilometer Array (SKA) radio telescope.  
We show that for axions lighter than a few $\times$ $10^{-13}$ eV a 100\,h SKA observation of the local dwarf galaxy Seg I would probe parameter space not constrained  
by stellar cooling and cosmological observations,  with sensitivity several orders of magnitude
better than the planned dedicated axion dark matter search experiments.
We quantify the uncertainties in the SKA sensitivity projections due to two effects that enhance the radio flux:
the presence of turbulent magnetic fields inside the galaxy, and the Bose enhancement of the DM decays to axions, where the latter, in particular, warrants further study. 
}


\begin{document}
\maketitle

\section{Introduction}

Pseudo-Nambu-Goldstone bosons are the low energy remnants of any
 spontaneously broken global symmetry, and are thus ubiquitous in beyond the Standard Model (BSM) theories. In modern parlance they are referred to as axion-like-particles (ALPs)~\cite{Kim:2008hd, Marsh:2015xka, DiLuzio:2020wdo}, the terminology that we will also adopt, often shortening it to simply ``axions''. 
 It is quite possible that there is a cosmic population of axions, signaling 
 spontaneous symmetry breaking that occurred in the early Universe~\cite{1992SvJNP..55.1063B}. 
These axions can be non-relativistic, constituting cold dark matter (DM)~\cite{Preskill:1982cy, Abbott:1982af, Dine:1982ah}, 
or relativistic, giving rise to a cosmic axion background (CAB) 
in the form of a dark radiation \cite{Conlon:2013isa,Marsh:2013opc,Dror:2021nyr,Langhoff:2022bij}. 

Axions in the mass range where these are good cold DM candidates are mostly searched through their couplings to photons: for $m_a$ below 1 $\mu$eV with lumped-circuits such as ABRACADABRA~\cite{Ouellet:2018beu, Ouellet:2019tlz} and DMRadio~\cite{Silva-Feaver:2016qhh, DMRadio:2022pkf, Brouwer:2022bwo}, for $m_a$ of 1$-$50 $\mu$eV with  haloscopes such as ADMX~\cite{ADMX:2003rdr, ADMX:2018gho, ADMX:2019uok}, 
HAYSTAC~\cite{PhysRevD.96.123008, HAYSTAC:2018rwy} 
and CAST-CAPP~\cite{Adair:2022rtw}, and for $m_a$ of $50\,\mu\,\text{eV}-1\,\text{meV}$ with a planned dielectric haloscope MADMAX~\cite{Caldwell:2016dcw, Millar:2016cjp}. These experiments can also search for relativistic cosmic axions. If the axions are produced in the early universe, the bounds on $\Delta N_{\rm eff}$~\cite{Planck:2018vyg,Baumann:2016wac}  restrict the CAB  energy density to be $\Omega_{\rm CAB}<\Omega_{\rm CMB}$ (for other cosmological and astrophysical bounds see~\cite{Dror:2021nyr, Eby:2021ece}, including bounds on axions from moduli decays~\cite{Conlon:2013txa, Moroi:2018vci, Higaki:2013qka, Jaeckel:2021ert, Jaeckel:2021gah}, and from evaporations of primordial black holes~\cite{Schiavone:2021imu}). For relativistic axions produced from late decays of DM the cosmological constraints are relaxed, and can lead to observable signals in terrestrial experiments, with ADMX (HAYSTAC) already constraining DM decay times into light axions to be several orders of magnitude longer than the age of the Universe ($t_U$)  for DM mass $m_{\chi}$ 
in the range 5$-$7 $\mu$eV (above a few tens of $\mu$eV to $\sim 100$ $\mu$eV)~\cite{Dror:2021nyr, Cui:2017ytb, Gu:2021lni}.

In the present manuscript we explore the potential of the upcoming radio telescope Square Kilometre Array (SKA)~\cite{Braun:2015B3, SKA}  to search for relativistic axions from DM decays, which subsequently converted to photons in the magnetic field of an astrophysical object such as the dwarf spheroidal galaxy (dSph). The wide range of frequencies ($\sim$50 MHz - $\sim$50 GHz) and  large effective area 
covered by the SKA relative to 
existing radio telescopes~\cite{SKA}, leads to high projected sensitivities for $m_{\chi}\sim10^{-7}$ to 
$\sim 10^{-2}$ eV. The projected sensitivity obtained with just  $\sim$100 hours of observation of 
local dSphs is typically already well above the reach of ADMX, HAYSTAC and DMRadio. 
Similar studies of SKA reach, but for axion cold dark matter with $m_a \gtrsim 10^{-8}$ eV were performed in~\cite{Kelley:2017vaa} using the Galactic center, in~\cite{Sigl:2017sew} using nebulae and other compact objects, in~\cite{Caputo:2018ljp} using dSphs, in~\cite{Battye:2019aco} using neutron star, the Galactic center and Galaxy cluster and in~\cite{Wang:2021hfb} using Globular cluster respectively as sources of radio signals as well as in~\cite{Ayad:2020fzc} where multiple different sources were used. In contrast, the relativistic axions probed by SKA are much lighter,  $m_a\lesssim10^{-12}$ eV, when constraints on axion-photon couplings~\cite{CAST:2017uph,Payez:2014xsa,Mirizzi:2009nq} are taken into account.  The signal from 7keV DM decaying to relativistic axions that subsequently convert to photons in magnetic field of galaxy clusters, was explored in~\cite{Cicoli:2014bfa} as a solution for the $3.5$ keV line anomaly. 

In  this manuscript  we focus on the local ultra-faint dSph galaxy Seg I as a useful target for SKA observations, 
mainly due to its high mass-to-light ratio and low star formation 
rate~\cite{Geha:2008zr, Simon_2011},  both of 
which suppress the relevance of highly uncertain contributions of various astrophysical processes when estimating the SKA reach. 
Observations of Seg I are additionally motivated by its proximity to the Earth and its high DM content, as can be inferred
from the observed mass-to-light ratio within its half-light radius. However, one can trivially adapt the analysis presented below
to other galactic systems as well.

The paper is structured as follows. 
In section~\ref{sec:axion_production} 
we discuss the production mechanism of relativistic axions from the decays of 
DM inside a dSph galaxy such as Seg I, as well 
as from decays of extragalactic and cosmological DM distributions, paying special attention to the effect of stimulated emission. 
In section~\ref{sec:conversion_probability} we review the phenomenon of axions converting into photons in a regular magnetic field of a dSph.
In section~\ref{sec:results} we 
present the projections for the SKA sensitivity to
the axion and DM parameters, 
with the effect of a turbulent magnetic field discussed in section~\ref{sec:turbulent_conversion}, 
and the effect of the Bose enhancement  in section~\ref{sec:proj:stim}.
We draw our conclusions in section~\ref{sec:conclusion}, while appendix \ref{sec:app:stimulated} contains further details on the calculation of stimulated emission.

\section{Axions from decaying dark matter}\label{sec:axion_production}

We assume that axions, $a$, are produced in the two body decays, $\chi\to aa$, where $\chi$ 
is a scalar DM~\cite{Dror:2021nyr, Cicoli:2014bfa, Cui:2017ytb}. 
If axions are much lighter than DM,  $m_a\ll m_{\chi}$, 
then the $\chi \to aa$ decays result in a population of highly relativistic axions. The axion energy spectrum receives two distinct contributions:  i) the Cosmic Axion Background (CAB) from  extragalactic DM decays, where the contributions from different epochs in the evolution of the universe result in a continuous axion energy spectrum,  and ii) from DM gravitationally bound inside various galactic structures,   
including dSphs, resulting in monoenergetic axions.

\subsection{Cosmic axion background from dark matter decays}
The differential number density of cosmic axions, ${dn_a}/{d\omega}$, 
produced from extragalactic DM decays throughout the evolution of the universe, is given by~\cite{Dror:2021nyr}, 
\begin{equation}
\left.\frac{dn_a}{d\omega}\right\vert_{\rm CAB} = \int_{t = 0}^{t_U} dt \hspace{1mm} {\hat{a}^3} \hspace{1mm} \Gamma_{\chi} e^{-\Gamma_{\chi} t} \hspace{1mm} \frac{\rho_{\rm DM}(t)}{m_{\chi}} \hspace{1mm} 
\left.\frac{dN_a}{d\omega'}\right\vert_{\omega' = \omega / \hat{a}},  
\label{eq:dna_dW}
\end{equation}
where $\omega$ is the axion energy at the present time of the Universe $t_U$, $\rho_{\rm DM} (t)$ is the DM energy density at time $t$, and $\Gamma_{\chi}=1/\tau_{\chi}$ 
is the DM decay rate. 
The initial axion energy spectrum at time $t$, ${dN_a}/{d\omega'}$,  is redshifted due to the expansion of the universe, with the present day axion energy given by $\omega = \hat a \omega' $, where the scale factor $\hat a$ is normalized such that $\hat a_0 = 1$ at present time.\footnote{We use $\hat a$ for the scale factor so that it is not confused with the axion field, $a$.}   In \eqref{eq:dna_dW} we assumed that axions remain relativistic throughout, which is an excellent approximation for the DM and axion mass ranges we will use in the numerical analysis below. 

\begin{figure*}[t]
\centering
\includegraphics[width=7cm]{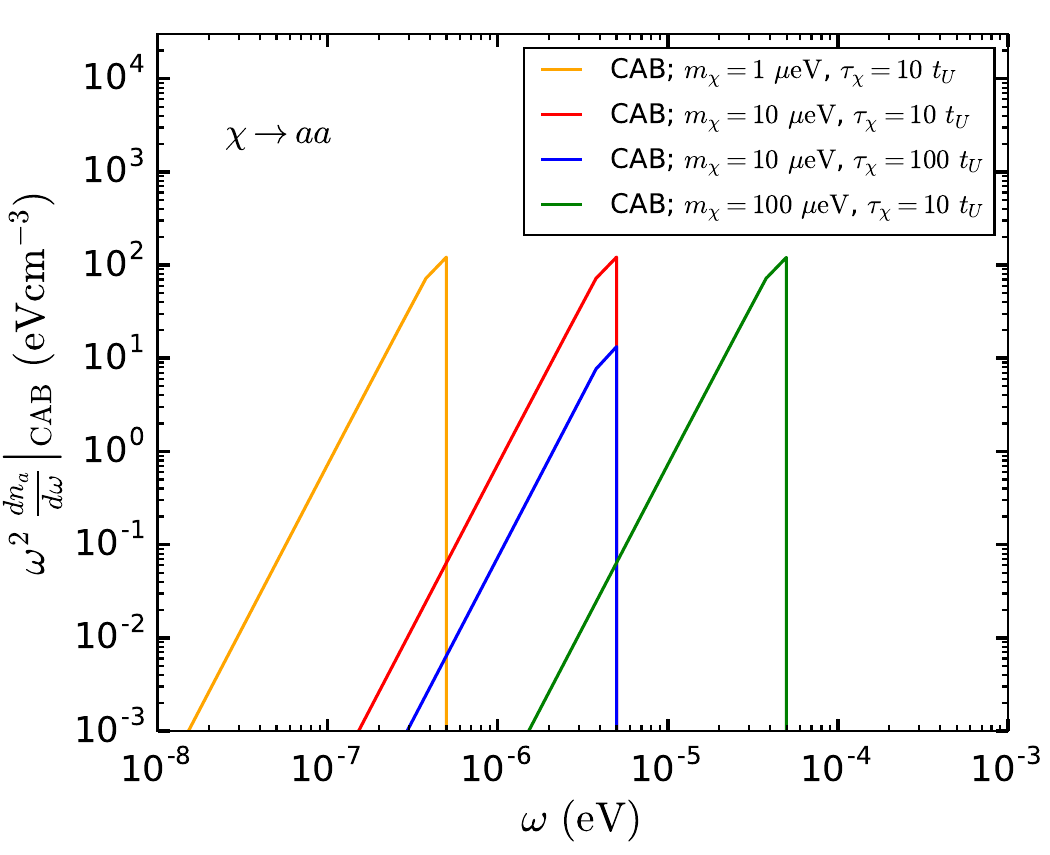} \\
\includegraphics[width=7cm]{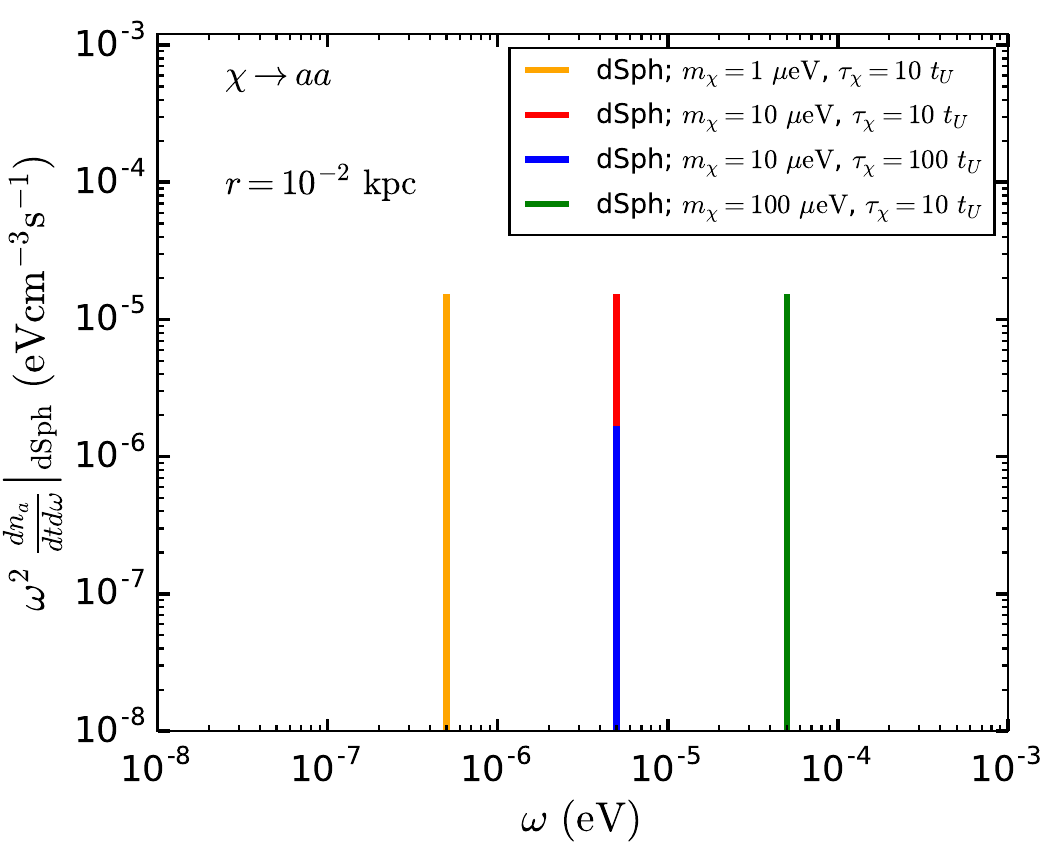}
\includegraphics[width=7cm]{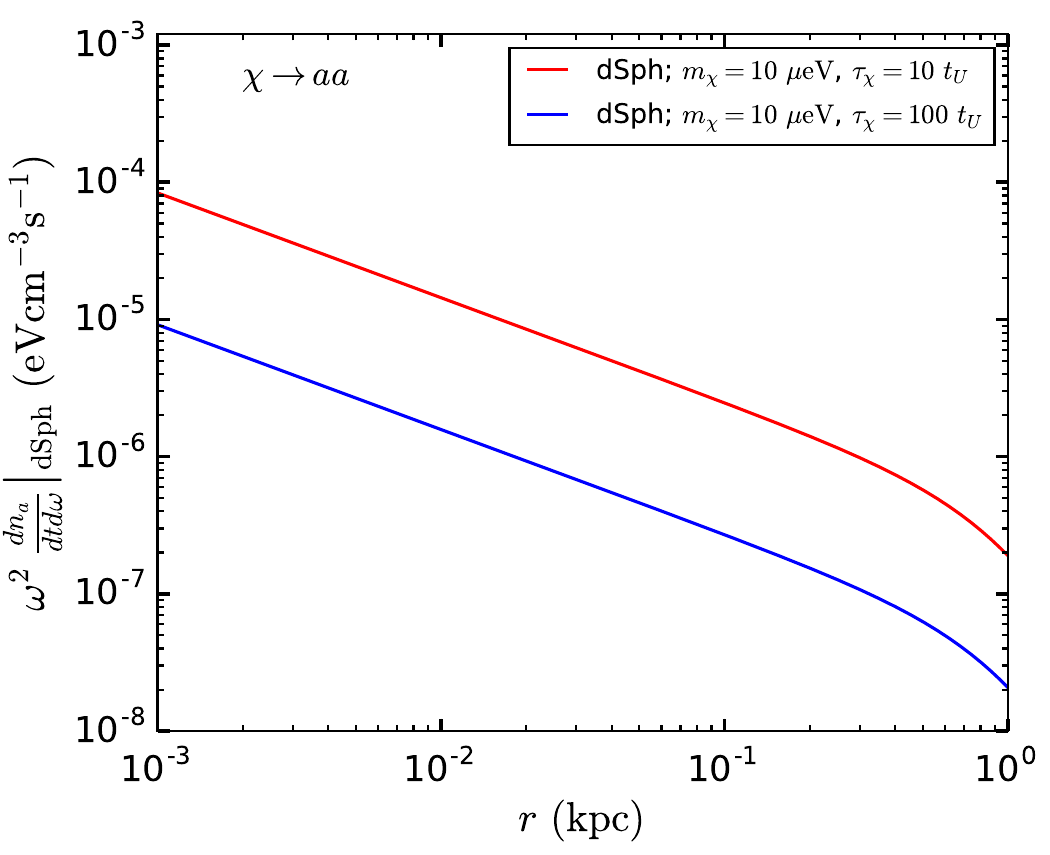}
\caption{{\it Top panel:} Present day CAB spectra  from extragalactic $\chi \rightarrow a a$ decays, as functions of axion energy $\omega$,
assuming the standard $\Lambda$CDM  cosmological history. 
Color coding for each choice of DM mass, $m_{\chi}$, and DM decay lifetime, $\tau_{\chi}$, 
is given in the legend. 
{\it Bottom left panel:} Differential axion flux generated from $\chi \rightarrow a a$ decays at $r = 10$\,pc radial distance from the center of Seg I dSph. 
{\it Bottom right panel:} 
DM induced axion flux inside Seg I dSph at peak axion energy, $\omega=m_\chi/2$, cf. discussion below Eq. \eqref{eq:dna_dw_dsph}, as a function of the radial distance $r$ from dSph center, for a fixed $m_{\chi}=10\mu\text{eV}$ and two values of $\tau_{\chi}=\{10,100\}t_U$. All plotted spectra ignore Bose enhancements of DM decays, the effect of which is discussed in Section \ref{sec:stimulated_emission}. 
}
\label{fig:CAB}
\end{figure*}

For two-body $\chi \rightarrow a a$ decays\footnote{We assume that $\chi \to a a$ is the dominant decay mode.  The UV physics that results in $a$ couplings to photons can also generate DM--photon couplings, and $\chi \rightarrow \gamma \gamma$ decays. While the relative size of $\chi \to a a$  and $\chi \to \gamma \gamma$ branching ratios is model dependent, typically one does have $Br(\chi \to a a)\gg Br(\chi \to \gamma\gamma)$ in large regions of parameter space (see also the discussion in~\cite{Calmet:2020rpx}).} 
the initial axion spectrum is $
{dN_a}/{d\omega'} = 2\delta(\omega' - m_{\chi}/2)$. Due to the cosmological redshift this results  in a  continuous present day axion spectrum with maximal energy $\omega=m_\chi/2$, i.e., the isotropic and homogeneous {\em  cosmic axion background (CAB)}, 
\begin{equation}
\left.\frac{dn_a}{d\omega}\right\vert_{\rm CAB} = 4 \frac{  \rho_{\rm DM,0}}{m_{\chi}^2}  \hspace{1mm} 
\frac{\exp\big({-\frac{1} \tau_{\chi}} t\big|_{\hat a=2 \omega / m_{\chi}} \big)}{\tau_{\chi} H\big|_{\hat a= 2 \omega / m_\chi}} \hspace{1mm} \Theta(m_{\chi}/2 - \omega).
\label{eq:dna_dW_new}
\end{equation}
Here, $\Theta$ is the Heaviside function, $\rho_{\rm DM,0}$ the present day DM energy density, while the Hubble rate $H$ and the time of DM decay, $t$,  
are evaluated at $\hat a=2\omega/m_{\chi}$. Fig.~\ref{fig:CAB} (top panel) shows the CAB spectra~\eqref{eq:dna_dW_new} for  $m_\chi=\{1,10,100\}\mu$eV and $\tau_\chi=\{10,100\}t_U$, where $t_U \simeq 13.8$ Gyr is the age of the universe. The CAB in the energy range $\omega\sim10^{-7}-10^{-4}$\,eV results, via axion-photon conversions in astrophysical magnetic fields, 
 in a radio signal in a frequency range $\nu = \omega / 2\pi\sim {\mathcal O}(10\text{s\,MHz})-{\mathcal O}(10\text{s\,GHz})$, which can be probed by radio telescopes (see the discussion below). 
To have an observable signal in SKA we therefore require in the numerical analysis for DM mass to be in the range $m_\chi\sim10^{-7}-10^{-4}$\,eV, see also Fig.~\ref{fig:CAB} (top panel). 

The amount of CAB sourced by the extragalactic DM decays is constrained by the observations of the CMB temperature, CMB polarization data and matter power spectra (while primordial CAB is bounded also by $\Delta N_{\rm eff}$).  Combining Dark Energy Survey, Planck-2018, supernovae type Ia and baryonic acoustic oscillation observations, Ref.~\cite{DES:2020mpv} finds no evidence  for dark-radiation \cite{Bringmann:2018jpr}. When interpreted in terms of DM decaying to dark radiation this leads to a bound $\tau_\chi \gtrsim 3.6 t_U$. In the numerical analyses below, we always choose $\tau_\chi$ benchmark values such that this bound is satisfied.

\subsection{Contributions from galactic DM decays}
\label{sec:dSph_contribution}

In addition to the homogeneous and isotropic CAB from decays of extragalactic DM, there is also a flux of relativistic axions from decays of galactic DM. This galactic component has directional variation. The largest axion flux originates in regions with the largest density of DM -- the galactic center and dSphs. We will be most interested in the latter, because of the reduced astrophysics backgrounds. 
The axion flux generated from DM decays 
per unit time and per unit volume of dSph galaxy is given by~\cite{Dror:2021nyr, Cicoli:2014bfa}, 
\begin{equation}
\left.\frac{dn_a}{dt d\omega}\right\vert_{\rm dSph} (r) = \frac{1}{m_{\chi} \tau_{\chi}} \frac{dN_a}{d\omega} \hspace{1mm} \rho^{\rm dSph}_{\rm DM} (r),
\label{eq:dna_dw_dsph}
\end{equation} 
where $\rho^{\rm dSph}_{\rm DM} (r)$ is the current DM density distribution of the dSph halo. For DM at rest, the axion energy spectrum is given by ${dN_a}/{d\omega} = 2 \delta(\omega - m_{\chi}/2)$. For dSph DM this is spread by 
$\sim 10^{-5}$, which is the velocity dispersion of DM bounded inside the dSph halo, and for observations on Earth additionally by the Earth's velocity, $\sim 10^{-3}$. Following Ref.~\cite{Dror:2021nyr}, we include the two effects by 
taking the spectrum ${dN_a}/{d\omega}$ in Eq.~\eqref{eq:dna_dw_dsph} as seen by terrestrial experiments
to be a Gaussian centered 
at $\omega=m_{\chi}/2$ and a width of $10^{-3} m_{\chi}/2$. 

The energy spectrum and the spatial distribution of the axion flux from DM decays in dSph galaxy Seg I, Eq.~\eqref{eq:dna_dw_dsph},  are shown in the two lower panels in Fig.~\ref{fig:CAB}. For DM distribution, $\rho^{\rm dSph}_{\rm DM} (r)$, 
we use the generalized density profile with the median values of the profile parameters in Table 4 of Ref.~\cite{Geringer-Sameth:2014yza}, obtained from a fit to
the stellar-kinematic data.

\subsection{The effect of Bose enhancement in DM decays}
\label{sec:stimulated_emission}
If the occupation number $\hat f_a$ of axions in the galaxy is large, this leads to Bose enhanced stimulated DM decay rate\footnote{Since for long lived DM $n_a\ll n_\chi$  we can neglect the back-reaction due to the 
$aa\to \chi$ annihilations, an approximation that we expect to have only a small effect on our results.} 
\begin{equation}
\frac{d n_{\chi}}{dt} = -n_{\chi} \Gamma_{\rm eff} ,
\label{eq:dnx_dt}
\end{equation}
where 
\begin{equation}
\Gamma_{\rm eff} = \Gamma_{\chi} (1 + 2\hat f_{a}), 
\label{eq:Gamma_eff}
\end{equation}
is the Bose enhanced effective $\chi\to a a$ decay rate, while $\Gamma_{\chi}$ is the perturbative decay rate (for a related discussion of stimulated $a\to \gamma\gamma$ decays see~\cite{Caputo:2018vmy, Alonso-Alvarez:2019ssa, Ayad:2020fzc}). 
Axion density at a radial distance $r$ from the center of dSph is given by,
\begin{equation}
n_a(r) = \int d\Omega \int_{l.o.s.} ds \frac{\Gamma_{\rm eff} N_a n_\chi}{4\pi},
\label{eq:na_r}
\end{equation}
where $N_a$ is the multiplicity of the final state, i.e., $N_a$=2 for $\chi\to aa$ , and the integral is over solid angle and the line of sight as viewed from the position in dSph for which we calculate $n_a$. In the integral the DM density should be retarded. However, in practice the change in $n_\chi$ during the time it takes the axion to traverse dSph is small enough that the effect of retardation can be neglected. 

The axion occupation number in the dSph halo,  $\hat f_a(\omega)$ in Eq.~\eqref{eq:Gamma_eff}, we approximate with~\cite{Alonso-Alvarez:2019ssa} 
\begin{equation}
\hat f_a \left(\omega = m_{\chi}/2 \right) \sim \frac{n_a}{4\pi \omega^2 \Delta \omega /(2\pi)^3}= \frac{2 \pi^2}{\omega^2} \frac{n_a}{\Delta\omega} .
\label{eq:fa}
\end{equation} 
Here, $\Delta\omega$ is the spread in the axion energy due to DM velocity dispersion, $v_d$, in dSph,  $\Delta\omega = {v_d} (m_{\chi}/2)$. 
For Seg I dSph we take $v_d = 10$ km/s~\cite{Geringer-Sameth:2014yza}. 
Note that the dispersion uncertainty in the axion energy is always larger than the one from the uncertainty principle $\Delta \omega \Delta t\geq 1/2$, i.e., $\Delta \omega \geq \Gamma_\chi/2$, for phenomenologically interesting perturbative decay widths, $\Gamma_\chi\ll 1/t_U$.  Furthermore, the contribution from CAB to axion occupation number $\hat f_a \left(\omega = m_{\chi}/2 \right)$ in dSph is negligible, and was thus not included in Eq.~\eqref{eq:fa}. 

\begin{figure*}[t]
\centering
\includegraphics[width=8cm]{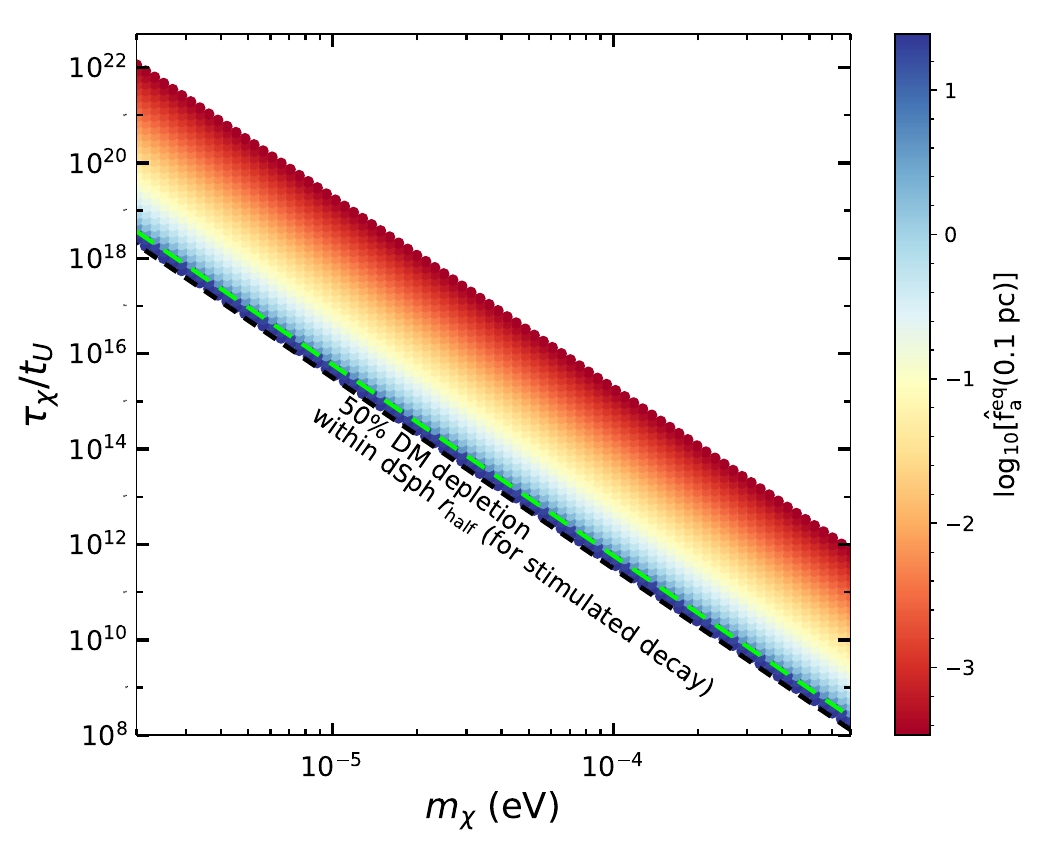}
\caption{Occupation number $\hat f_a$ (at a radius 0.1 pc inside the dSph)
as a function of $m_\chi$ and $\tau_\chi/t_U$ obtained by solving iteratively Eq.~\eqref{eq:na_r}, see Appendix \ref{sec:app:stimulated} for details. 
The black dashed line indicates the values of $\tau_{\chi}$ below which  our iterative procedure results in more than 50\% DM depletion within 
the half-light radius of the dSph. 
 Our iterative procedure is however reliable only above the green dashed line, which denotes the $\tau_{\chi}$  values 
(obtained numerically) 
for which $\Delta t_a$ (the time it takes for $n_a$, at 0.1 pc, to change by a factor of 2) is equal to $\Delta t_{\rm dSph}$.}
\label{fig:fahat}
\end{figure*}

Since $\hat f_a$, and  consequently $\Gamma_{\rm eff}$, depend on $n_a$, cf. Eqs.~\eqref{eq:Gamma_eff} and \eqref{eq:fa}, this means that Eq.~\eqref{eq:na_r} is an integral equation for $n_a(r)$, and as such  is difficult to solve analytically. We provide instead an approximate numerical solution for the case of a slowly changing axion density, i.e., in the limit $\Delta t_a\gg \Delta t_{\rm dSph}$. 
Here, $\Delta t_{\rm dSph}$ is the typical time required for a relativistic axion to traverse the dSph, $\Delta t_{\rm dSph}\sim 1/1\text{kpc}\sim 3\cdot 10^3$ years, while $\Delta t_a$ is the typical time in which $n_a$ changes by an ${\mathcal O}(1)$ factor. Taking $\Delta n_a/\Delta t_a\simeq - dn_\chi/dt$, with $\Delta n_a\sim n_a$ gives
\beq
\label{eq:Deltata:estimate}
\Delta t_a\simeq \frac{\Delta n_a}{n_\chi \Gamma_\chi (1+2 \hat f_a)}\sim \frac{m_\chi^4 v_d}{32 \pi^2 \rho_\chi \Gamma_\chi},
\eeq
where in the last relation we assumed $\hat f_a\gg 1$, with $\hat f_a$ given by \eqref{eq:fa} for $\omega=m_\chi/2$, and replaced $n_\chi=\rho_\chi/m_\chi$. Note that $n_a$ drops out from the last estimate for $\Delta t_a$ in \eqref{eq:Deltata:estimate} since $\hat f_a\propto n_a$. Numerically, 
\beq
\Delta t_a\simeq 10^{-27} \biggr(\frac{m_\chi}{1\,\mu\text{eV}}\biggr)^4\biggr(\frac{v_d}{10^{-5}}\biggr)\biggr(\frac{0.1 M_{\odot}/\text{pc}^3}{\rho_\chi}\biggr)\tau_\chi.
\eeq
Requiring $\Delta t_a\gg \Delta t_{\rm dSph}$ then gives 
\beq
\label{eq:tauchi/tU}
\tau_\chi/t_U\gtrsim 10^{20} \Big(1\,\mu\text{eV}/m_\chi\Big)^4,
\eeq
for the parameter region where the approximation of a slowly changing axion density is valid.

In Fig. \ref{fig:fahat} we show the iterative numerical solution for $n_a(r)$ that follows from Eq.~\eqref{eq:na_r}, after it is converted to the occupation number $\hat f_a$ at $r=0.1\,$pc using Eq. \eqref{eq:fa}. The iterative solution relies on the assumption of a slowly changing axion density, see details in Appendix  \ref{sec:app:stimulated}. It is valid above the green dashed line  in Fig.~\ref{fig:fahat}, obtained numerically by requiring that  the change from $n_a$ to $2n_a$ takes longer than $\Delta t_{\rm dSph}$ throughout the iterative procedure. This bound coincides well with the analytical estimate, Eq. \eqref{eq:tauchi/tU}. Fig. \ref{fig:fahat} shows that in the parameter region where the iterative procedure  is reliable, the effect of the stimulated emission is either negligible (for $\tau_\chi$ well above  the bound \eqref{eq:tauchi/tU}, in which case $\hat f_a\ll 1$), or only modestly important (for $\tau_\chi$ close to the bound \eqref{eq:tauchi/tU}, in which case $\hat f_a\sim {\mathcal O}(10)$). 

Extending the numerical analysis to even smaller values of $\tau_\chi$ we find a runaway behavior for the $n_a$ density so that quickly ${\mathcal O}(1)$ of DM decays within the age of the dSph (the 50\% depletion is denoted with a black dashed line in Fig. \ref{fig:fahat}). Since this occurs in the parameter region where our numerical approach is no longer reliable we cannot draw strong conclusions. On the one hand the observed runaway behavior could be a result of our numerical approach and the effect of stimulated emission remains modest also for smaller values of $\tau_\chi$. However, it is equally plausible that the runaway is physical, in which case the DM model we consider  faces a coincidence problem -- as we will see below,  in this case an observable signal in SKA would require parameters close to the ${\mathcal O}(1)$ DM depletion line, raising a ``why now'' question?

Resolving this issue would require a more detailed numerical study, with a finely discretized description of dSph, potentially in 3D. Such a study goes beyond the scope of this paper. Instead, we show in Section \ref{sec:results} two sets of projections for the SKA sensitivity: first ignoring the effect of stimulated emission, and then including its effect, but limiting the analysis to the parameter region where our approximation is valid.  In order to obtain the SKA reach, however, we have to first discuss the conversion of axions into photons in the galactic magnetic fields, which then leads to the observable signal in SKA.

\section{Axion-photon conversion}
\label{sec:conversion_probability}

Axions couple to photons through a dimension 5 operator~\cite{Raffelt:1987im}
\begin{equation}
\mathcal{L}_{a\gamma} = -\frac{1}{4} g_{a\gamma\gamma} F_{\mu\nu} \tilde{F}^{\mu\nu} a.
\end{equation}
Typically, we expect $g_{a\gamma\gamma}\simeq \alpha_{\rm em}/(4\pi f_a)$, where $f_a$ is the axion decay constant. A value of $g_{a\gamma\gamma}$ much smaller than this naive estimate would  require a fine tuned cancellation between the UV and the nonperturbative QCD  contributions in the IR, which we assume not to be the case.  

Due to the axion-photon interaction $\mathcal{L}_{a\gamma}$ the axion flux converts to photon flux in the magnetic field of a galaxy.  We start by reviewing in Sec. \ref{sec:single:domain} the results for axion-photon conversion in a region with homogeneous magnetic field, and then give numerical estimates for the resulting radio flux from Seg I in Sec. \ref{sec:radio_flux}.

\subsection{Conversions in a single magnetic domain}
\label{sec:single:domain}

The probability for an axion with mass $m_a$ and energy $\omega$ to convert to a photon, after traversing a coherent magnetic field domain of length $L_{\rm domain}$, is given by~\cite{Grossman:2002by, Schiavone:2021imu, Carenza:2021alz}, 
\begin{equation}
P_{a\gamma} = {\rm sin^2} \left(2 \theta \right) \hspace{1mm} {\rm sin^2} \left(\frac{\Delta_{\rm osc} L_{\rm domain}}{2} \right).
\label{eq:P_agamma}
\end{equation}
Both the mixing angle, 
\begin{equation}
{\rm tan}(2 \theta) = \frac{2 \Delta_{a\gamma}}{\Delta_{\parallel} - \Delta_a},
\end{equation}
and the oscillation frequency, 
\begin{equation}
\Delta_{\rm osc} = \sqrt{(\Delta_{\parallel} - \Delta_a)^2 + 4 \Delta_{a\gamma}^2} \hspace{1mm} ,
\end{equation}
depend on the product of the transverse component of the magnetic field, $B_{\bot}$, and the axion-photon coupling constant, 
\begin{equation}
\Delta_{a\gamma} = \frac{1}{2} g_{a\gamma\gamma} B_{\bot} 
                \simeq 1.52 \times 10^{-3} {\rm kpc^{-1}} \times \left( \frac{g_{a\gamma\gamma}}{10^{-12} {\rm GeV^{-1}}} \right) 
                 \left( \frac{B_{\bot}}{\mu G} \right) \hspace{1mm} ,
\end{equation}
where in the numerical example we chose a representative value of $g_{a\gamma\gamma}$ that can be constrained by SKA for large range of axion masses, and the value of a magnetic field typical for dSph. 

\begin{figure*}[t]
\centering
\includegraphics[width=7cm]{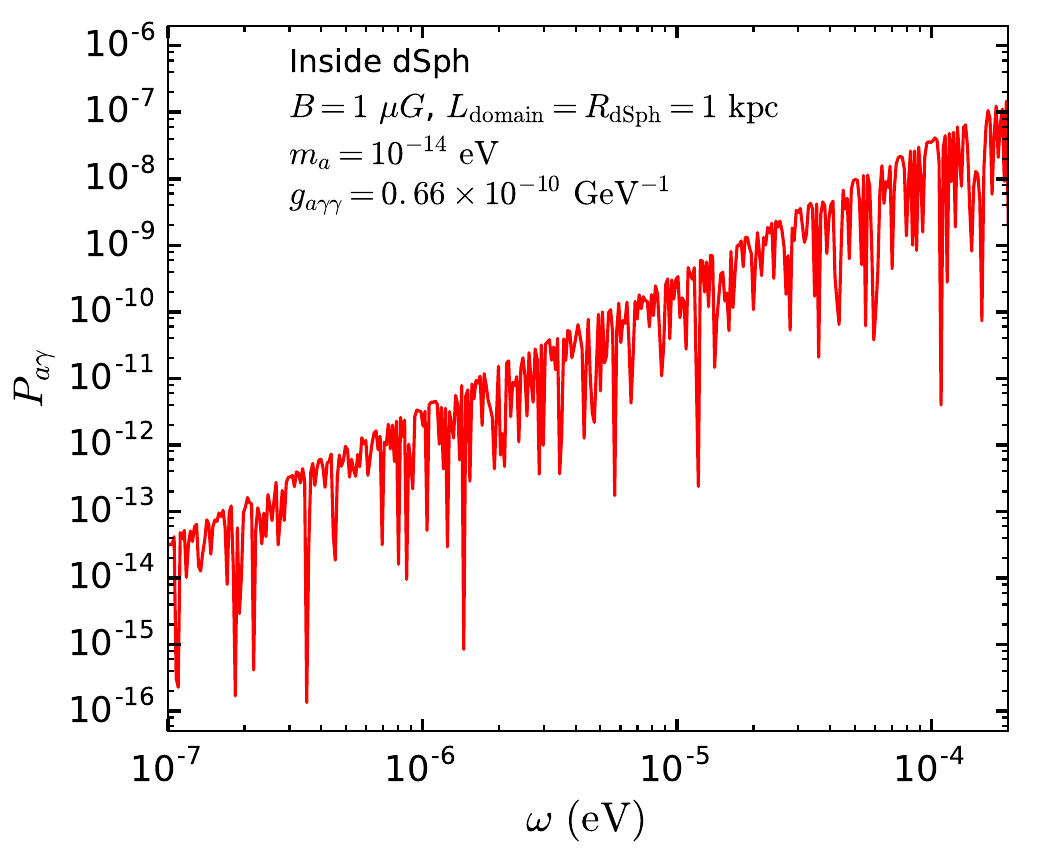}~~~~
\includegraphics[width=7cm]{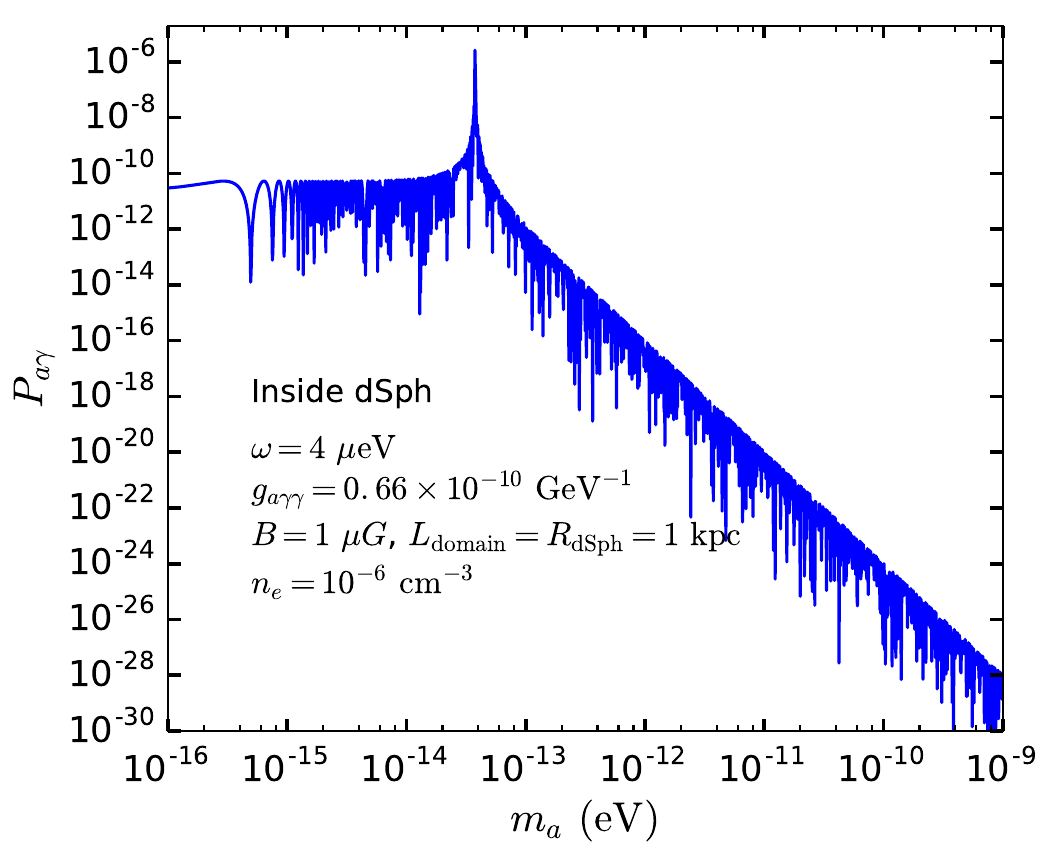}
\caption{Axion$\rightarrow$photon conversion probability $P_{a\gamma}$ as a function of the axion energy $\omega$ (left panel) and axion mass $m_a$ (right panel), for a single magnetic domain assuming a homogeneous magnetic field with a value typical for magnetic fields in a dSph (values as denoted in the legends). 
The sharp peak 
at $m_a\simeq 4\times 10^{-14}$ eV indicates resonant conversion.}
\label{fig:P_agamma}
\end{figure*}

The oscillation frequency corresponding to axion-photon conversions should be compared with the mismatch between the oscillations induced by the axion mass corrections
\begin{equation}
\Delta_{a} = - \frac{m_a^2}{2 \omega}  
           \simeq - 7.8 \times 10^5 \left( \frac{m_a}{10^{-13} {\rm eV}} \right)^2 
           \left( \frac{\omega}{\mu {\rm eV}} \right)^{-1} \hspace{1mm} {\rm kpc^{-1}},
\end{equation}
and the plasma induced effects
\begin{equation}
\Delta_{\parallel} = \Delta_{\rm pl} + 3.5 \Delta_{\rm QED},
\end{equation}
where
\begin{align}
\Delta_{\rm pl} &= - \frac{\omega^2_{\rm pl}}{2 \omega} \simeq - 1.1 \times 10^5 \left( \frac{\omega}{\mu {\rm eV}} \right)^{-1} 
                \left( \frac{n_e}{10^{-6} {\rm cm^{-3}}} \right) \hspace{1mm} {\rm kpc^{-1}}, 
\label{eq:del_pl}                
\\
\Delta_{\rm QED} &\simeq 4.1 \times 10^{-24} \left( \frac{\omega}{\mu {\rm eV}} \right) 
\left( \frac{B_{\bot}}{\mu G} \right)^2 \hspace{1mm} {\rm kpc^{-1}} , 
\label{eq:del_QED}
\end{align}
with $\omega_{\rm pl}\simeq 3.7 \times 10^{-11}\text{eV}\times  \sqrt{{n_e}/{\rm cm^{-3}}}$  the plasma frequency, and 
$n_e$ the electron density of the plasma in dSph. For $n_e \sim 10^{-6}\,\text{cm}^{-3}$, which is a typical value for dSphs~\cite{Colafrancesco:2006he, McDaniel:2017ppt}, the $\Delta_{\rm pl}$ and $\Delta_{a}$ are of the same order for axion with a mass $m_a \sim 4 \times 10^{-14}$ eV.

 Fig.~\ref{fig:P_agamma} shows the dependence of the axion$\rightarrow$photon conversion probability $P_{a\gamma}$ inside a dSph, Eq.~\eqref{eq:P_agamma},  on either the axion energy $\omega$ (left panel) or on the axion mass $m_a$ (right panel). 
In the evaluation the magnetic field was set to $B_{\bot} = B = 1\,\mu G$, a typical  value for dSph galaxies such as Seg I~\cite{Colafrancesco:2006he, McDaniel:2017ppt, Natarajan:2015hma} (in Sec.~\ref{sec:radio_flux} we will also show results for a more conservative value, $B = 0.1\,\mu G$). The size of the domain was taken to coincide with 
the dSph radius, $L_{\rm domain}=R_{\rm dSph}=1$\,kpc~\cite{Arshakian:2008cx}, while the axion-photon coupling $g_{a\gamma\gamma} = g^{\rm SE}_{a\gamma\gamma}=0.66 \cdot 10^{-10}$\,GeV${}^{-1}$ was chosen such that it saturates the stellar emission bounds for $m_a < 10^{-9}\,$eV~\cite{Dror:2021nyr, CAST:2017uph}. 
The conversion probability $P_{a\gamma}$   shows a sharp peak  at $m_a \simeq 4 \cdot 10^{-14}\,\text{eV}$ (cf. Fig.~\ref{fig:P_agamma} right), for which $\Delta_{\parallel} \simeq \Delta_a$, resulting in a resonant conversion. For heavier axion masses the conversion probability rapidly decreases as $P_{a\gamma} \propto m_a^{-4}$.

\subsection{Radio flux from conversions in a dwarf spheroidal galaxy}
\label{sec:radio_flux}

Axions that convert to photons in the magnetic field of the dSph galaxy give rise to a potentially observable signal in the radio wave frequency range. 
The resulting 
radio flux density as observed on Earth, is given by 
 \begin{equation}
S = \left.S\right\vert_{\rm CAB} + \left.S\right\vert_{\rm dSph},
\label{eq:radio_flux}
\end{equation}
where $S\big|_{\rm CAB}$ is the component of the flux from the CAB axions converting to photons while traversing the dSph, while $S\big|_{\rm dSph}$ is due to converted axions from DM decays in the dSph itself~\cite{Cicoli:2014bfa}.

The frequency distribution of the 
flux density due to CAB axions after traversing a domain of length $L_{\rm domain}$ that contains homogeneous magnetic field is given by~\cite{Conlon:2013txa}, 
\begin{equation}
\left.S (\nu) \right\vert_{\rm CAB} = \frac{P_{a\gamma}(\omega)}{L_{\rm domain}} \hspace{1mm}
\times \frac{1}{4 \pi \nu} \int_{\Omega} \int_{\rm l.o.s} d\Omega \hspace{1mm} ds \hspace{1mm} \omega^2 \left.\frac{dn_a}{d\omega}\right\vert_{\rm CAB}
\hspace{2mm}
\text{(in the units of $\rm erg\; cm^{-2} s^{-1} Hz^{-1}$)} , 
\label{eq:radio_flux_CAB}
\end{equation}
where $\nu = \omega / 2\pi$ is the frequency of the detected photon. 
The integration over dSph volume is in terms of a line of sight (l.o.s) coordinate $s$ and a solid angle  $\Omega$. 
The CAB number density ${dn_a}/{d\omega}\big|_{\rm CAB}$ 
is given in~\eqref{eq:dna_dW}, while 
$P_{a\gamma}/L_{\rm domain}$ is the rate per unit time at which axions convert inside a dSph~\cite{Conlon:2013txa}. Here, $L_{\rm domain}$ is the time it takes for an ultrarelativistic axion to traverse the dSph, while the conversion probability $P_{a\gamma}$ is given by Eqs.~\eqref{eq:P_agamma}--\eqref{eq:del_QED}. The total flux density on Earth is then obtained by summing different contributions along the line of sight.

Similarly, the radio flux density from axions that were emitted in DM decays within dSph halo and then convert to photons in the domain of length $L_{\rm domain}$ is given by~\cite{Caputo:2018ljp, Cicoli:2014bfa}, 
\begin{equation}
\left.S (\nu) \right\vert_{\rm dSph} = P_{a\gamma}(\omega) \hspace{1mm} 
\times \frac{1}{4 \pi \nu} \int_{\Omega} \int_{\rm l.o.s} d\Omega \hspace{1mm} ds \hspace{1mm} \omega^2 \left.\frac{dn_a}{dt d\omega}\right\vert_{\rm dSph} 
\hspace{2mm}
\text{(in the units of $\rm erg\; cm^{-2} s^{-1} Hz^{-1} $)} , 
\label{eq:radio_flux_dsph}
\end{equation}
where $\left.{dn_a}/{dt d\omega}\right\vert_{\rm dSph}$ is given in Eq.~\eqref{eq:dna_dw_dsph}, and $P_{a\gamma}$ by Eqs.~\eqref{eq:P_agamma}--\eqref{eq:del_QED}. As for CAB component, the total flux density observed on Earth is given by summing different contributions along the line of sight. For SKA projections in Section \ref{sec:results} we will consider two cases: that dSph has a completely coherent magnetic field, as well as the possibility that the magnetic field has a turbulent component. 
In the numerical estimates of the 
flux density $S$ 
we use 
the conventional units of Jy, 
where 1 Jy = $10^{-23}$ $\rm erg$ $\rm cm^{-2} s^{-1} Hz^{-1}$. 

\begin{figure*}[t]
\centering
\includegraphics[width=7cm]{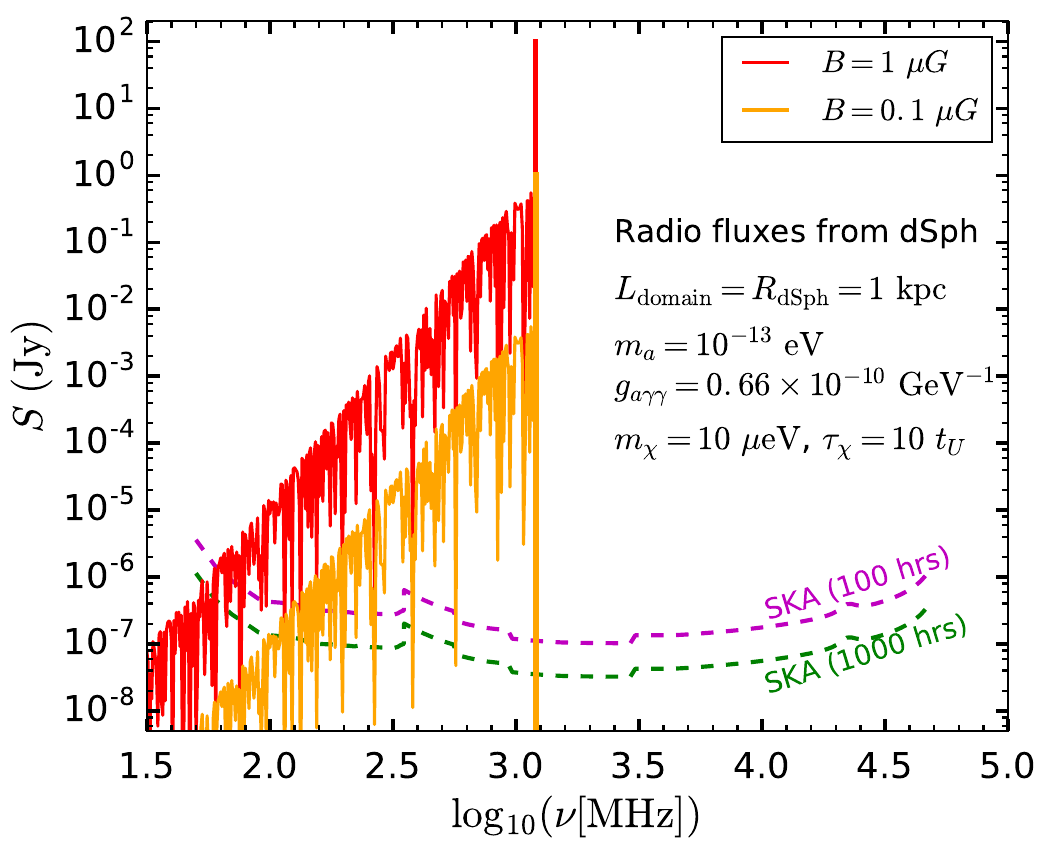}
\caption{The red (orange) line shows the  
radio flux density from Seg I  dSph from conversion of  relativistic axions (produced in $\chi\to aa$ decays),  assuming $B = 1 \mu G$ $(0.1\mu G)$ homogeneous magnetic field inside Seg I (the other parameters are as listed in the legend, and we ignore the effect of stimulated emission). The broad continuum is due to the CAB, while the monochromatic line at the high frequency cut-off is from  
the DM decays in the dSph halo. 
The magenta (green) dashed line shows the projected SKA sensitivity for 100 (1000) hours of observations.}
\label{fig:flux}
\end{figure*}

Before moving to the SKA projections, it is insightful to calculate the axion conversion 
radio flux density $S$ that can be expected from the ultra-faint dSph Seg I. As a representative example we take $m_{\chi} = 10\,\mu$eV, $\tau_{\chi} = 10 t_U$, $m_a = 10^{-13}$\,eV 
and $g_{a\gamma\gamma} = 0.66 \times 10^{-10}$ $\rm GeV^{-1}$, and consider the case of a homogeneous  magnetic field $B = 1(0.1)\,\mu\text{G}$ inside the dSph, i.e., we take $L_{\rm domain}=R_{\rm dSph}=1$ kpc, and $n_e = 10^{-6}\;\rm cm^{-3}$  (in line with Refs.~\cite{Arshakian:2008cx, McDaniel:2017ppt, Natarajan:2015hma, Regis:2014koa}), and use the same DM profile for Seg I as in Section~\ref{sec:dSph_contribution}. 
The predicted 
flux density is shown with red (orange) line in Fig.~\ref{fig:flux}. It consists of a sharp monochromatic line at $\nu=(m_\chi/2) (1/2 \pi)$, due to the DM decays inside dSph, and a tail at lower frequencies, due to conversions of CAB axions. The flux density due to DM decays in the dSph gives the dominant contribution to the integrated radio flux density, a consequence of a much higher DM density in the dSph halo. Note that the expected 
radio flux density drops with decreased magnetic field, in line with the expected decrease in the conversion probability, cf. Eq.~\eqref{eq:P_agamma}. 
Fig.~\ref{fig:flux} also shows the expected sensitivities of the SKA radio telescope after 100 (magenta dashed line) and 1000 hours of observations (green dashed line).  Further details about these projections are given in the next section.

\section{Projected SKA sensitivity}
\label{sec:results}
Next, we discuss the SKA sensitivity to axion parameters, $m_a$, $g_{a\gamma\gamma}$, and DM parameters, $m_\chi, \tau_\chi$, assuming that SKA would observe Seg I for either 100 or 1000 hours. 

\subsection{Details of SKA: Sensitivities and field of view}
The Square Kilometer Array (SKA) is a next generation radio telescope array currently under construction. The SKA is expected to operate in a wide frequency spectrum stretching from roughly few tens of MHz to few tens of GHz. The SKA will also make use of various technologies to expand its field of view (FOV), thereby enabling it to instantaneously image a large portion of the sky~\cite{2008IAUS..248..164T,2012PhDT.......123B,2021ExA....51....1B}. The SKA will also allow us to digitally patch multiple FOVs to observe a wide region of interest in the sky~\cite{Chen:2021rea}. Similar technique of patching multiple FOVs to image a large portion of the sky at a time is also used by other telescopes like the Jansky Very Large Array~\cite{VLA, 1988A&A...202..316C}. These techniques, in turn, will enable the SKA array to attain an effective FOV ranging from 200 square degrees to 1 square degree, depending on the frequency of observation~\cite{2008IAUS..248..164T}. Thus apart from a small portion in the high frequency end of the spectrum, in the majority of the radio frequency spectrum, the entire emission region of dSph like Seg I can be visible within the effective FOV of the SKA array. Consequently, for simplicity, we consider that almost all the radio flux originating from the entire dSph emission region can be captured by SKA and will contribute to the detected signal.\\
\indent Furthermore, the SKA is also projected to be the most sensitive radio telescope till date. Since the telescope is yet to start operations, exact values of the rms surface brightness sensitivity of the SKA array across the frequency range, in the direction of the dSph, are not yet available. 
Using the presently accepted SKA baseline design, ref.~\cite{SKA} has provided an averaged $(T_{\rm sys} / A_{\rm eff})$ for SKA-Low and SKA-Mid arrays, which is averaged over all solid angles within 45 degrees of the zenith, as a function of the observing frequency. Here $A_{\rm eff}$ is the effective collecting area of the array and $T_{\rm sys}$ is the system temperature.
For this work, in order to present our bounds on the axion and DM parameter spaces, we have considered this averaged $(T_{\rm sys} / A_{\rm eff})$ to estimate the SKA sensitivity. The expected rms surface brightness sensitivity or equivalently the rms noise level of both the SKA-Low and SKA-Mid arrays, $N_{\rm rms}$, are then obtained as~\cite{Caputo:2018ljp, Ghara:2015mab}
\begin{equation}
    N_{\rm rms} = 2 \frac{T_{\rm sys}}{A_{\rm eff} \sqrt{\Delta \nu \;t_{\rm obs}}}
\end{equation}
where $\Delta \nu$ is the instantaneous bandwidth and $t_{\rm obs}$ is the observation time. 
For calculating the instantaneous bandwidth $\Delta \nu$ we follow the presently accepted baseline design~\cite{SKA}.
The expected rms surface brightness sensitivity for two values of $t_{\rm obs}$ - 100 hours and 1000 hours - are shown in Fig.~\ref{fig:flux} by the dashed magenta and green lines respectively.
To simplify our calculation we assume that the array sensitivity of SKA for an enlarged FOV is the same as those shown in Fig.~\ref{fig:flux}. In order to get a more accurate sensitivity, a detailed calculation is required that will depend on the actual configuration of the SKA when it is fully operational, which at present is beyond the scope of this work.

We then define the SKA projected reach as in~\cite{Kar:2019cqo}: the radio signal is deemed to be detectable if it is 3 times larger than the rms noise level $(N_{\rm rms})$ in at least one frequency bin in the SKA frequency range, $\sim$50 MHz to $\sim$50 GHz~\cite{SKA}.

The projections have two main uncertainties: the knowledge of the magnetic field in Seg I, and whether or not the stimulated axion emission is  an important effect. To assess the first uncertainty we consider several values of the magnetic field, to start with assuming  in Section \ref{sec:coh:magn:field} that the magnetic field is to a good approximation homogeneous, and then including the effect of a turbulent component in Section \ref{sec:turbulent_conversion}, in both cases ignoring stimulated emissions. The projections for the parameter region for which our calculation of stimulated emission is valid are discussed in Section \ref{sec:proj:stim}.

\subsection{Projections assuming a coherent magnetic field}
\label{sec:coh:magn:field}

\begin{figure*}[t]
\centering
\includegraphics[width=7cm]{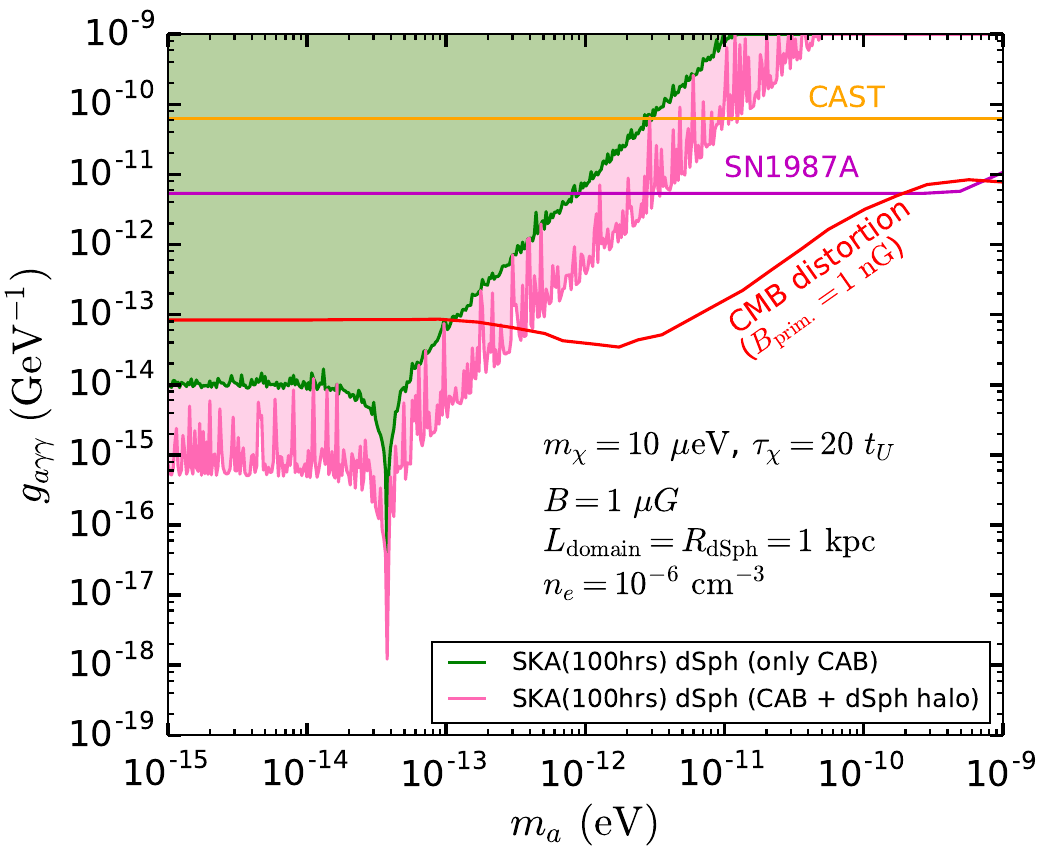}
\includegraphics[width=7cm]{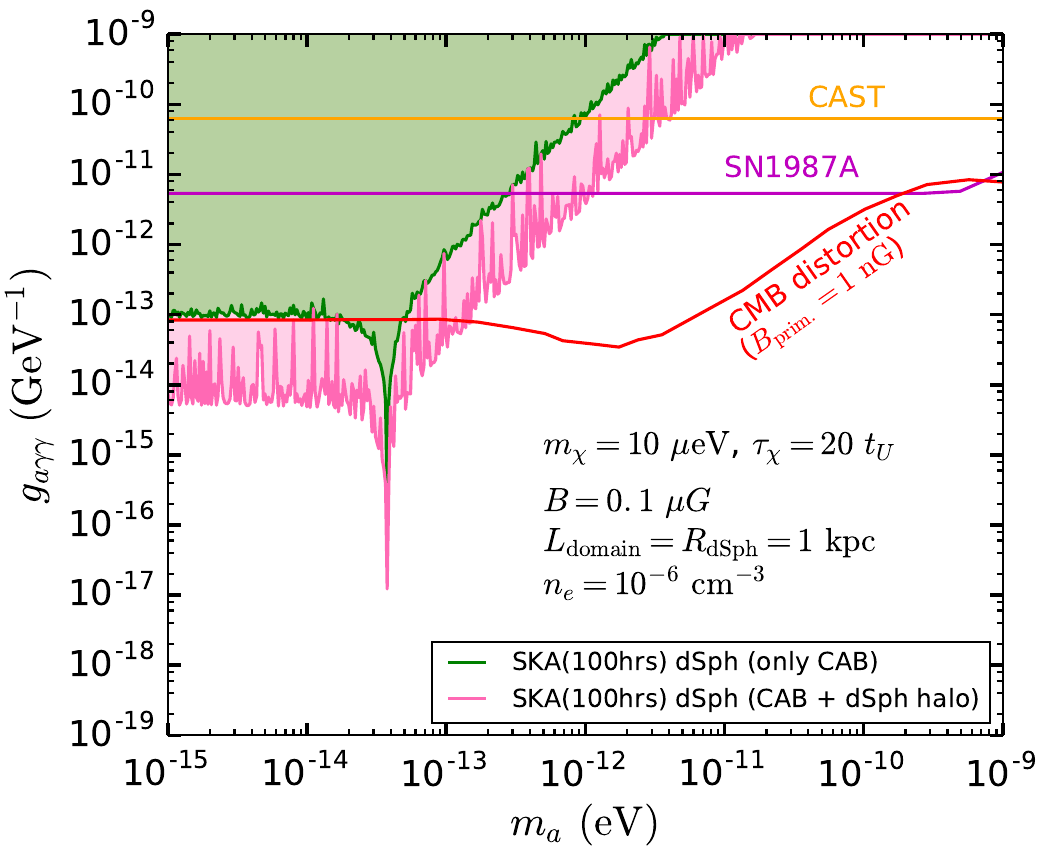}
\caption{Regions in pink show  the projected SKA reach in the $m_a - g_{a\gamma\gamma}$ plane, assuming either $B = 1\,\mu\text{G}$ regular 
magnetic field  (left panel) or $B = 0.1\,\mu\text{G}$ (right panel), and 100 hours of observation time in the direction of dSph Seg I    
(green region shows the reach from the CAB contribution to the radio flux). 
The DM and astrophysical parameters are as listed in the panels. Solid lines indicate present upper-limits on $g_{a\gamma\gamma}$ from CAST (orange), SN 1987A (magenta) and CMB distortions (red). 
}
\label{fig:limit_g_ma}
\end{figure*}

\begin{figure*}[t]
\centering
\includegraphics[width=7cm]{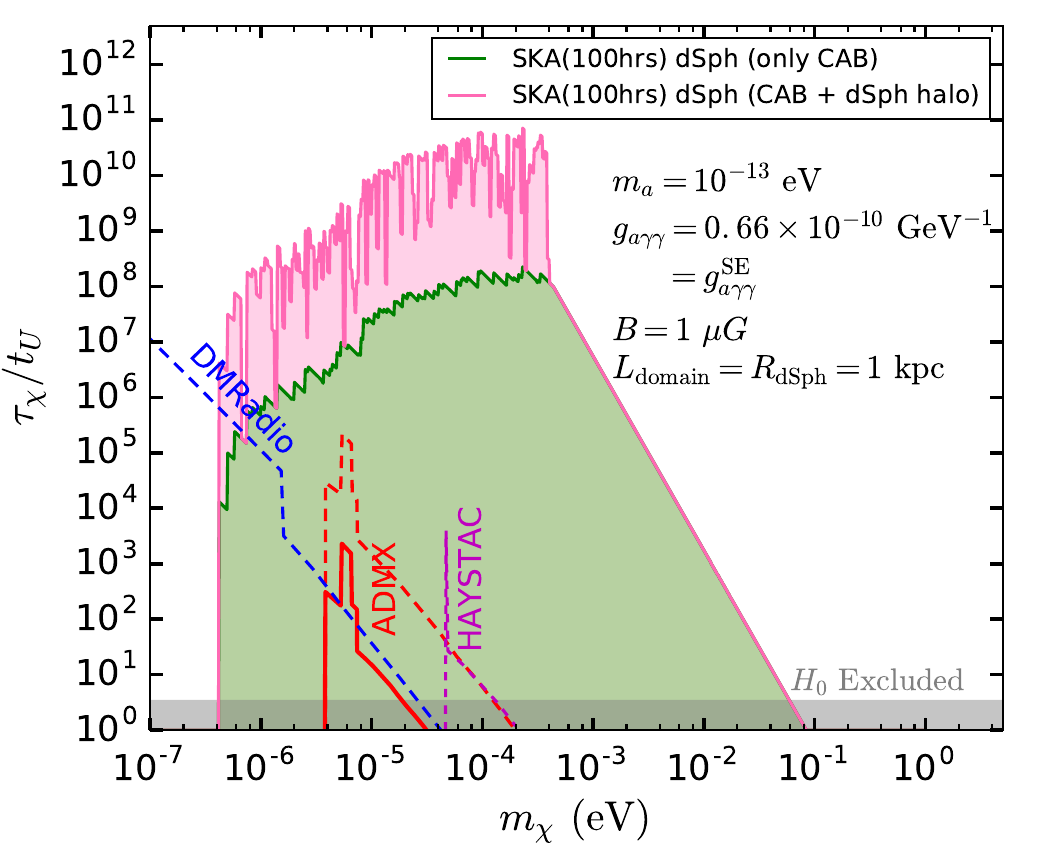}\hspace{0mm}
\includegraphics[width=7cm]{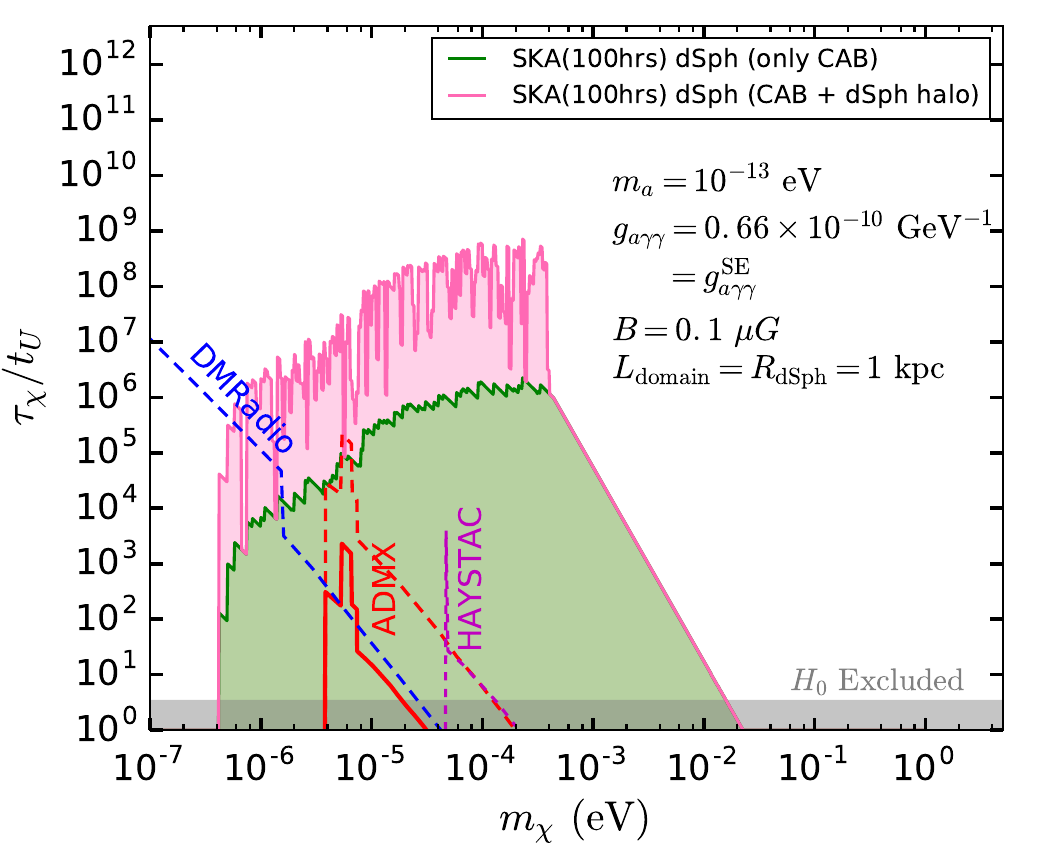}
\caption{Same as Fig. \ref{fig:limit_g_ma}, but for the projected 100h SKA reach in the DM parameter space. 
The solid red line denotes the present bound (lower-limit) from the ADMX experiment, while dashed lines give future sensitivities 
of ADMX (red), HAYSTAC (magenta) and DMRadio (blue). 
The gray shaded  region is excluded by $H_0$ measurements, see main text for further details. 
}
\label{fig:limit_tau_mdm}
\end{figure*}

Figs.~\ref{fig:limit_g_ma} and~\ref{fig:limit_tau_mdm} show in pink the projected SKA reach in the $m_a - g_{a\gamma\gamma}$   and $m_\chi-\tau_\chi$ plane, respectively, assuming that the magnetic field in Seg I is homogeneous, with its flux density equal to either $B = 1 \mu G$ (left panels, a more typical value) or  $B = 0.1 \mu G$ (right panels, a conservative value). We use the same values of $L_{\rm domain}=R_{\rm dSph}=1$\,kpc and $n_e= 10^{-6}\,\text{cm}^{-3}$ 
as in Fig.~\ref{fig:P_agamma}. 
The green shaded regions in Figs.~\ref{fig:limit_g_ma} and~\ref{fig:limit_tau_mdm} indicate the SKA reach that would be obtained from the subdominant CAB contribution to the radio flux.

In the example in Fig.~\ref{fig:limit_g_ma} we set DM lifetime to $\tau_{\chi} = 20$ $t_U$ and DM mass to $m_{\chi} = 10 \mu$eV. For larger DM lifetimes the reach in $g_{a\gamma\gamma}$ drops as $\sim 1/\sqrt{\tau_\chi}$, while for heavier DM masses the region of SKA sensitivity moves to higher $m_a$. 
The sharp increase in the SKA sensitivity at $m_a\simeq 4\times 10^{-14}$eV 
is the result of a resonant conversion, 
where the resonant 
$m_a$ depends on the assumed value of electron number density as $\sim \sqrt{n_e}$, see Section~\ref{sec:single:domain}. 
For $m_a\gtrsim$ few$\times 10^{-13}$\,eV the SKA sensitivity falls rapidly with the increasing $m_a$, following  the corresponding drop in the 
axion-photon conversion probability, c.f. Fig.~\ref{fig:P_agamma}.  For $m_a\lesssim$ few$\times 10^{-13}$\,eV, on the other hand, the projected SKA reach is well beyond the present bounds on axion couplings to photons: the present bound from 
CAST helioscope search for solar axions (orange line)~\cite{CAST:2017uph}, the bound due to non-observation of gamma-rays from conversions of SN 1987A emitted axions (magenta)~\cite{Payez:2014xsa}, and from bounds on CMB distortions that would be caused by conversion of CMB photons into axions (red)~\cite{Mirizzi:2009nq}. Note that the CMB distortions depend strongly on the value of the primordial magnetic flux density. For the exclusion line in Fig.~\ref{fig:limit_g_ma} we assume 
$B_{\rm prim.} = 1\,{\rm nG}$, comparable to
the current upper limits from CMB measurements~\cite{Planck:2015zrl}.

Fig.~\ref{fig:limit_tau_mdm} shows the projected SKA reach in the 
$m_{\chi} - \tau_{\chi}$ plane, setting the axion parameters to $m_a = 10^{-13}$ eV and $g_{a\gamma\gamma} = 0.66 \times 10^{-10} \rm GeV^{-1}$, and the same color coding as in Fig.~\ref{fig:limit_g_ma}.  For smaller values of $g_{a\gamma\gamma}$ the reach in $\tau_\chi$ drops as $1/g_{a\gamma\gamma}^2$. For heavier $m_a$ the parameter region probed by SKA would move to higher values of DM mass, with a diminishing reach, while for smaller $m_a$ the reach can be even much higher, cf. Fig.~\ref{fig:limit_g_ma}. The DM mass range probed also depends on the value of the $B$-field in the target dSph: for the examples shown in Fig.~\ref{fig:limit_tau_mdm}, SKA will probe a DM mass in the range  $4 \times 10^{-7}\;\rm eV$ to $\sim10^{-1}\;\rm eV$ (left panel) or 
to $\sim 10^{-2}\;\rm eV$ (right panel). 
Note that for axion parameters used in Fig.~\ref{fig:limit_tau_mdm}, DM lighter than $4 \times 10^{-7}\;\rm eV$ results in a conversion photon signal below the 
observable frequency range of the SKA, resulting in a sharp drop in the SKA sensitivity at low DM masses. At the upper end of the probed DM mass range, the monochromatic part of the signal is outside the SKA frequency range, and thus the signal in SKA  is  entirely due to the CAB contribution, leading to a slow drop in sensitivity for higher masses due to the falling axion flux (this is proportional to the average DM number density and thus $\propto 1/m_\chi$). Note that the SKA reach in the relevant DM mass range is well above the present exclusions: from the $H_0$ measurements (gray shaded area) \cite{DES:2020mpv}, from ADMX 
(solid red line)~\cite{Dror:2021nyr}, as well as future sensitivities from ADMX (dashed red), HAYSTAC (dashed magenta) and DMRadio (dashed blue), see Ref.~\cite{Dror:2021nyr} for further details on these projections.

\subsection{Turbulent magnetic field}
\label{sec:turbulent_conversion}

\begin{figure*}[t]
\centering
\includegraphics[width=6cm]{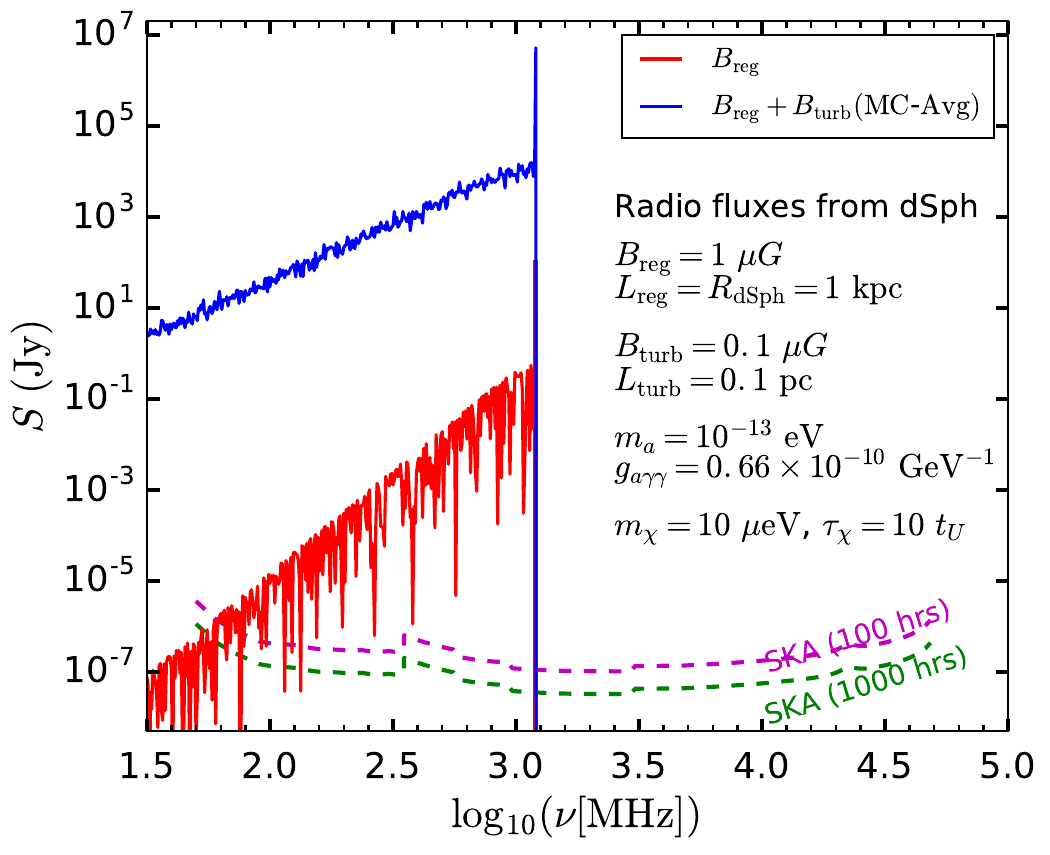} \\
\includegraphics[width=6cm]{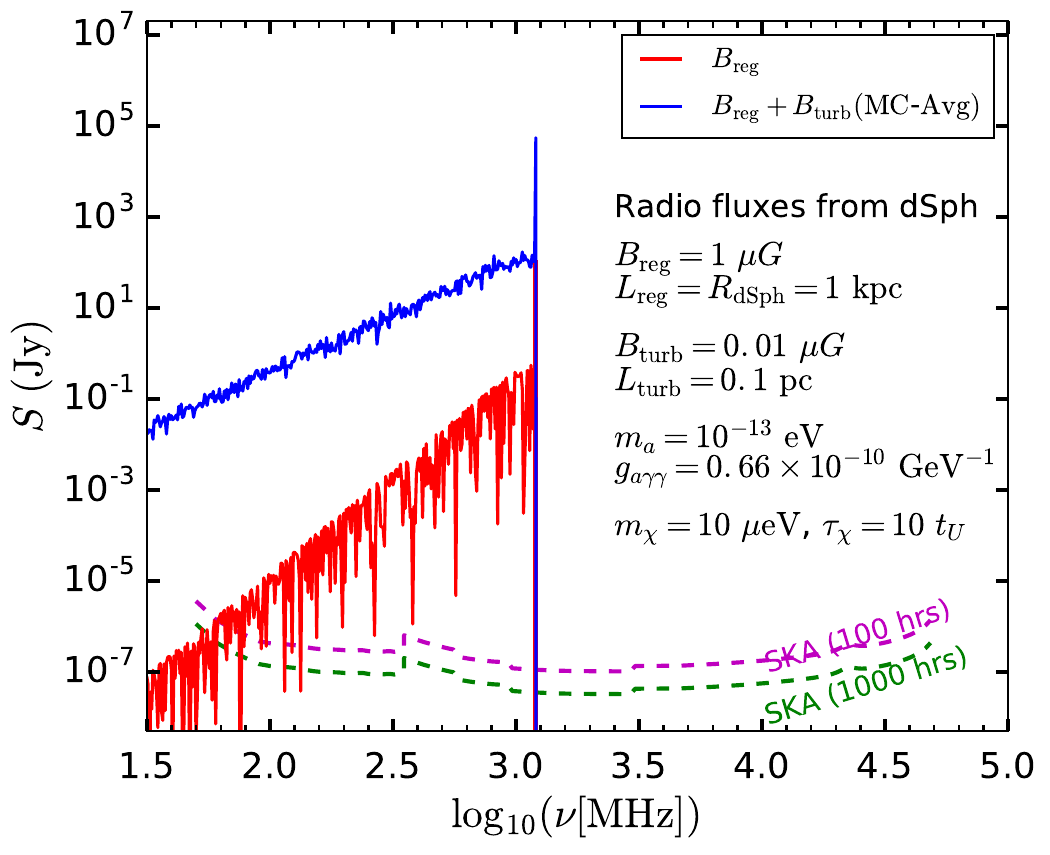}
\includegraphics[width=6cm]{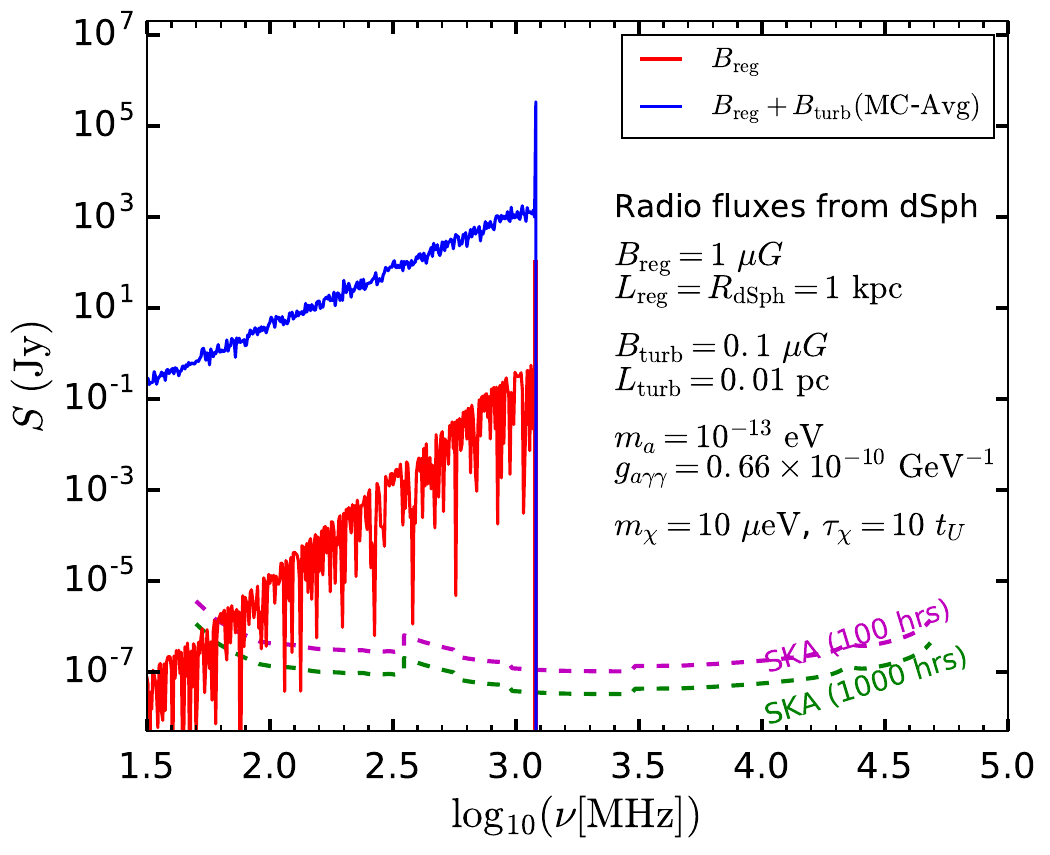}
\caption{The 
radio flux density $S$ from $\chi\to aa$ axions converted to photons in the magnetic field of Seg I, either assuming regular magnetic field (red curve) or in addition the turbulent component with the parameters listed in Eq. \eqref{eq:Bturb}: (a) -- top panel, (b) -- bottom left, (c) -- bottom right. The SKA sensitivity for 100 (1000) hours of observation is shown with magenta (green) dashed line. The axion and DM parameters are as listed in the panels. 
}
\label{fig:flux_Ptot}
\end{figure*}

So far we assumed that dSphs have only regular magnetic field with coherence length 
$\sim 1 \; \rm kpc$. Galaxies, including dSphs, can also host a turbulent 
magnetic field component with small coherence length, $L_{\rm turb}$, and root mean square (rms) amplitude, $B_{\rm turb}$~\cite{Caputo:2018ljp, Regis:2014koa,Beck_2011,Carenza:2021alz,Kachelriess:2021rzc}. As we show below,
the addition of such a small scale turbulent magnetic field increases the axion$\rightarrow$photon conversion probability $P_{a\gamma}$ \cite{Carenza:2021alz}, and consequently increases the SKA reach. 

In galaxies such as Milky Way and M31, $B_{\rm turb}$ is typically of the same order as 
the strength of the regular field, $B_{\rm reg}$,  while $L_{\rm turb}$ is two to three orders of magnitude smaller than the coherence length of the regular 
field, $L_{\rm reg}$~\cite{Carenza:2021alz, Conlon:2014xsa}. For dSphs the values of $B_{\rm turb}$ and $L_{\rm turb}$, on the other hand, are highly uncertain. 
For Seg I we explore three representative cases, 
\begin{subequations}
\label{eq:Bturb}
\begin{align}
\text{(a)~}& B_{\rm turb} = 0.1 \mu G, \qquad L_{\rm turb} = 0.1 \text{~pc}, 
\\
\text{(b)~} & B_{\rm turb} = 0.01 \mu G, \quad\,\,\, L_{\rm turb} = 0.1 \text{~pc}, 
\\
\text{(c)~} & B_{\rm turb} = 0.1 \mu G, \qquad L_{\rm turb} = 0.01 \text{~pc}, 
\end{align}
\end{subequations}
For regular magnetic field we take, as in Section  \ref{sec:coh:magn:field}, $B_{\rm reg} = 1\,\mu\text{G}$ and $L_{\rm reg} = R_{\rm dSph} = 1\,\text{kpc}$. 

Fig.~\ref{fig:flux_Ptot} shows the predicted 
radio flux density from axion conversions in Seg I, either assuming  just the regular magnetic field component (red lines) or in addition also the turbulent magnetic field component (blue lines), for the three representative cases in \eqref{eq:Bturb}. 
The radio flux density in the case of a turbulent magnetic field was obtained from the ``cell-model" of Ref.~\cite{Carenza:2021alz}, with the cell size set to $L_{\rm cell} = 2 L_{\rm turb}$. For each parameter set in \eqref{eq:Bturb} we performed a Monte Carlo (MC) simulation with 10 realizations of the turbulent field, divided into cells, calculated axion$\rightarrow$photon conversion probability $P_{a\gamma}$, and then averaged it over the 10 realizations.  In calculating the 
radio flux density both CAB and the dSph halo contributions were included, with the  axion and DM parameters set to $m_a=10^{-13}$ eV, $g_{a\gamma\gamma}=0.66\times 10^{-10}\text{\,GeV}^{-1}$, $m_\chi=10\mu$eV, $\tau_\chi=10t_U$. 

\begin{figure}[t]
    \centering
    \includegraphics[width=7cm]{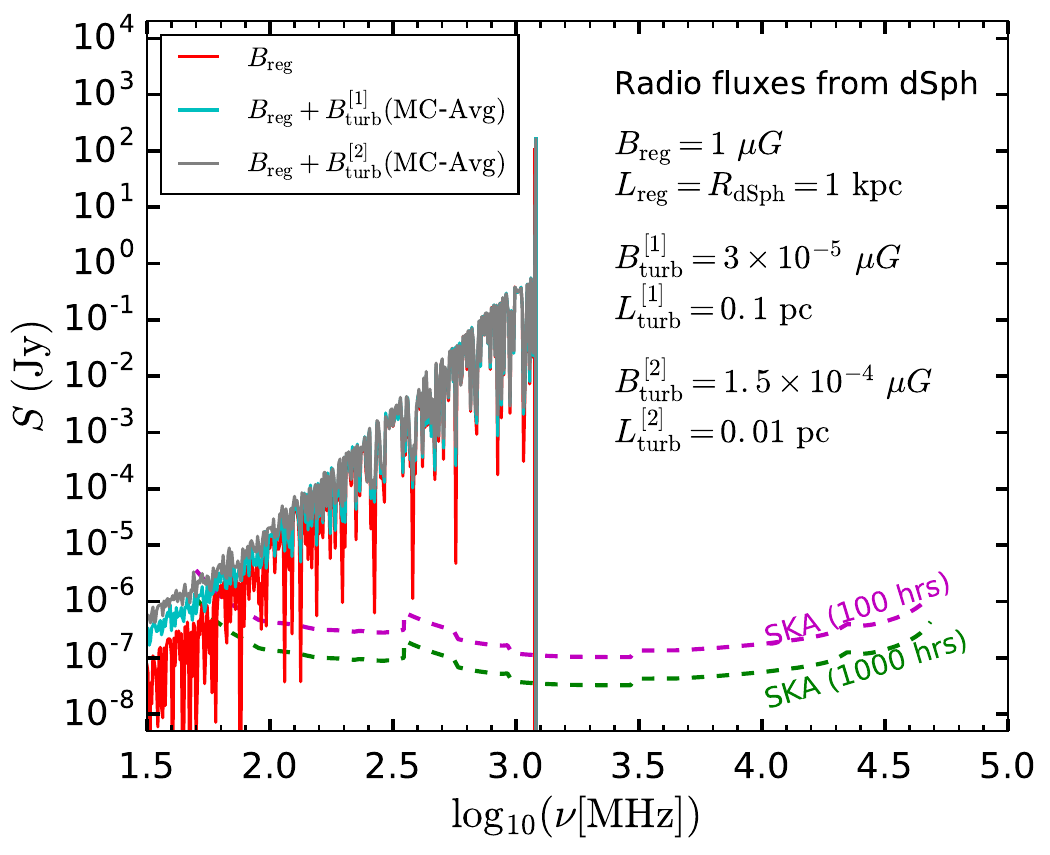}
    \caption{For small enough turbulent magnetic fields, e.g.,  $B_{\rm turb} = 3 \times 10^{-5}\;\mu$G for $L_{\rm turb} = 0.1\;$pc (cyan) and $B_{\rm turb} = 1.5 \times 10^{-4}\;\mu$G for $L_{\rm turb} = 0.01\;$pc (gray) the radio flux density 
    $S$ is indistinguishable from the case of a regular magnetic field (red). 
    The values of axion, DM, and regular magnetic field parameters, as well as the color coding for SKA sensitivities, are as in Fig.~\ref{fig:flux_Ptot}.
    }
    \label{fig:limit_flux}
\end{figure}

The enhancement of the radio flux density $S$ due to the turbulent magnetic field is several orders of magnitudes. It depends on both $B_{\rm turb}$ and $L_{\rm turb}$, where for chosen parameters $S$ increases if either  $B_{\rm turb}$ or $L_{\rm turb}$ increase, see Fig.~\ref{fig:flux_Ptot}. 
For small enough $B_{\rm turb}$ the axion-photon conversion probability becomes indistinguishable from having just the regular magnetic field, see the examples in Fig.~\ref{fig:limit_flux}. 
  The threshold value of $B_{\rm turb}$, below which the effect of turbulent magnetic field is negligible, depends on the value of $L_{\rm turb}$, and is for instance $B_{\rm turb} \simeq 3 \times 10^{-5}\, \mu\text{G}$ for $L_{\rm turb} = 0.1\,$pc (cyan line), and $B_{\rm turb} \simeq 1.5 \times 10^{-4}\; \mu G$ for  $L_{\rm turb} = 0.01\;$pc (grey), see Fig.~\ref{fig:limit_flux}, for the axion and DM parameter as in Fig.~\ref{fig:flux_Ptot}. 
The threshold value of $B_{\rm turb}$ 
also depends on $m_a$ and $m_{\chi}$. For example, for $m_a > 10^{-13}$ eV, 
the threshold value of $B_{\rm turb}$ is smaller than in the two examples.  
In all cases we checked that the threshold values of $B_{\rm turb}$ 
we obtain here are well below the value of 
$B_{\rm turb}$ expected in local dSphs such as Seg I, if one assumes 
similar scaling between $B_{\rm reg}$ and $B_{\rm turb}$ as is usually assumed 
for Milky Way.
The projected SKA sensitivities obtained 
in Section \ref{sec:coh:magn:field} can thus be viewed as 
fairly conservative (note though that these still neglect stimulated emission).

\begin{figure*}[t]
\centering
\includegraphics[width=6cm]{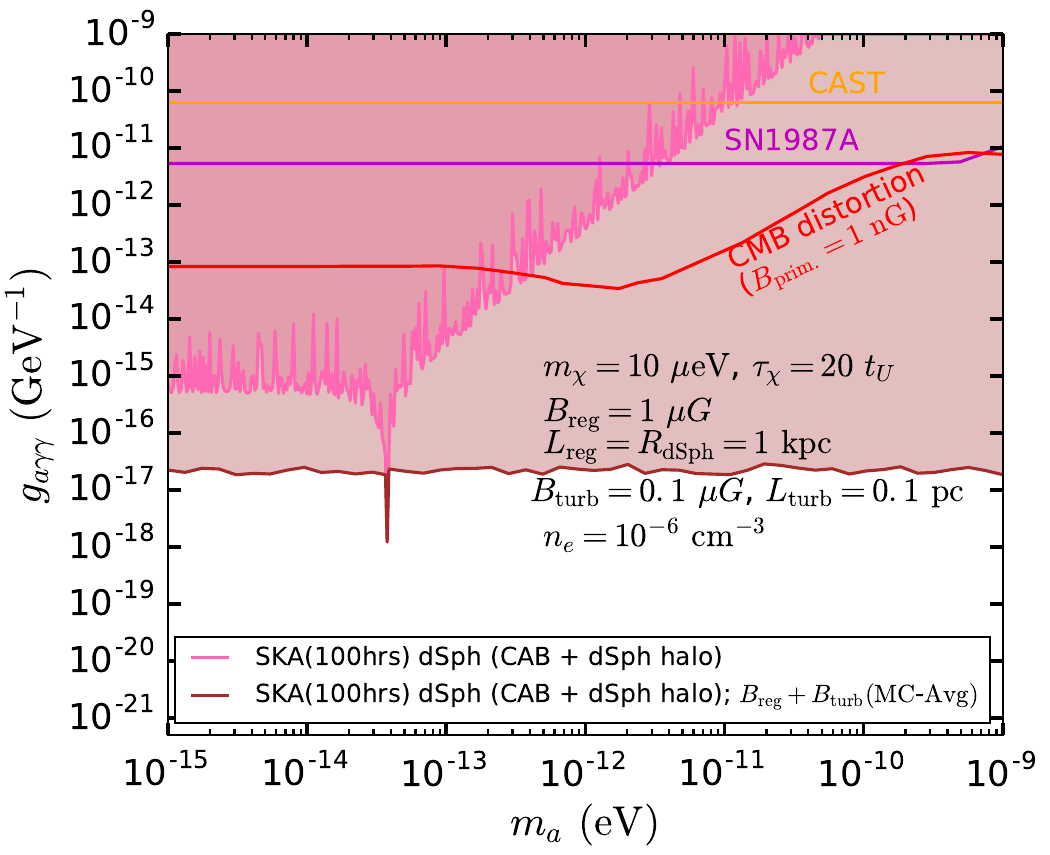}\\
\includegraphics[width=6cm]{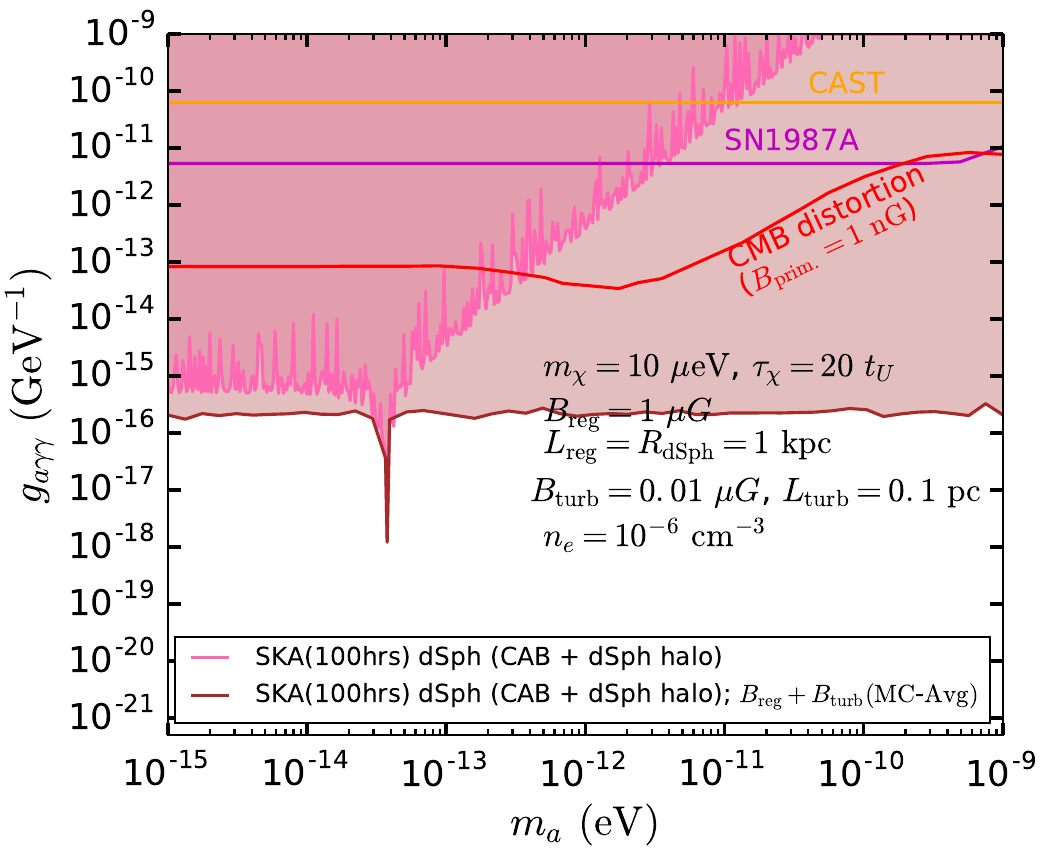}
\includegraphics[width=6cm]{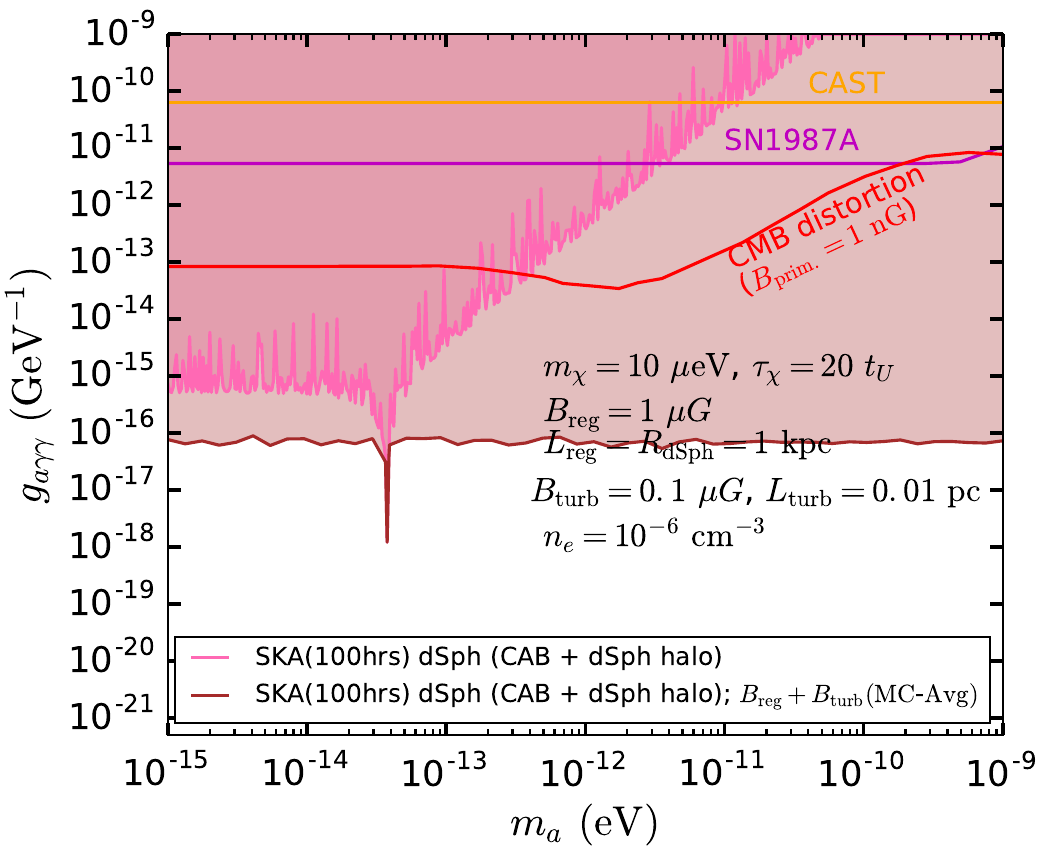}
\caption{The expected SKA sensitivity reach in the $m_a - g_{a\gamma\gamma}$ plane after 100 hours of Seg I observations, for either just the regular magnetic field (pink) or for, in addition, the turbulent component (brown) from Eq. \eqref{eq:Bturb}: (a) -- top panel, (b) -- bottom left, (c) -- bottom right. The DM and the regular magnetic field parameters are as in Fig.~\ref{fig:flux_Ptot}. 
The orange, magenta and red solid lines denote the existing bounds from CAST, SN 1987A and CMB distortions, respectively.
}
\label{fig:limit_g_ma_Ptot}
\end{figure*}

\begin{figure*}[t]
\centering
\includegraphics[width=6cm]{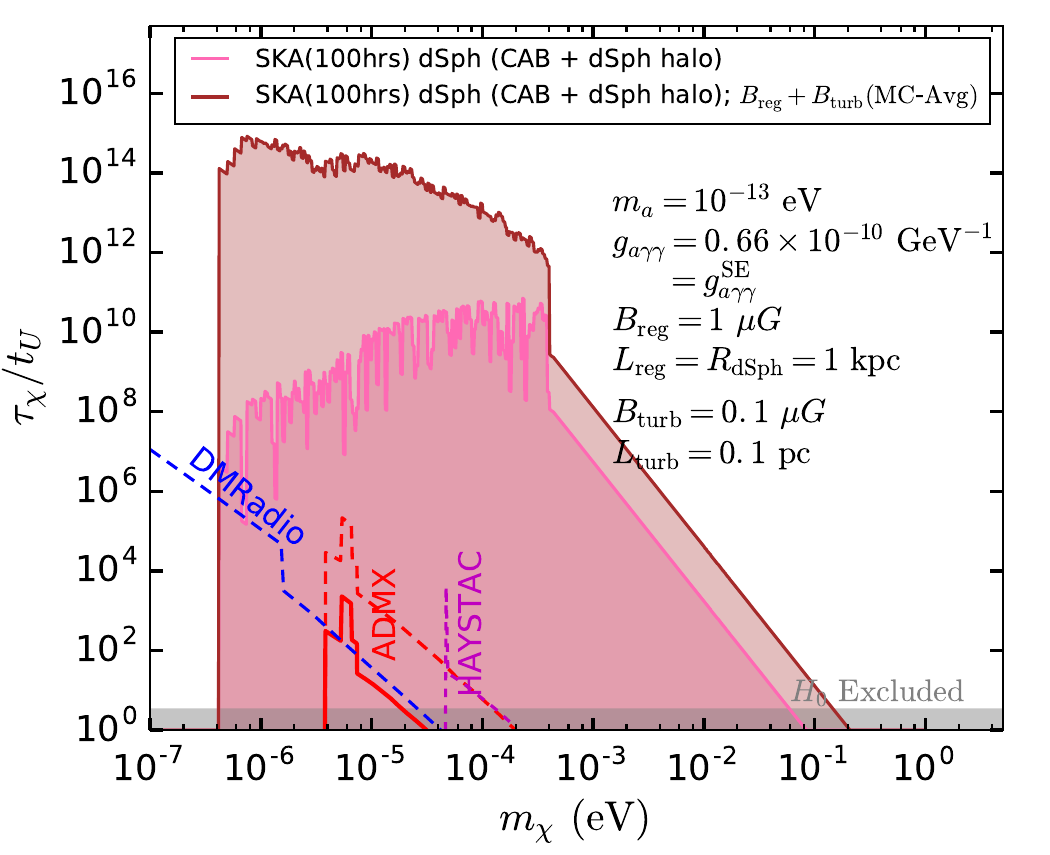} \\
\includegraphics[width=6cm]{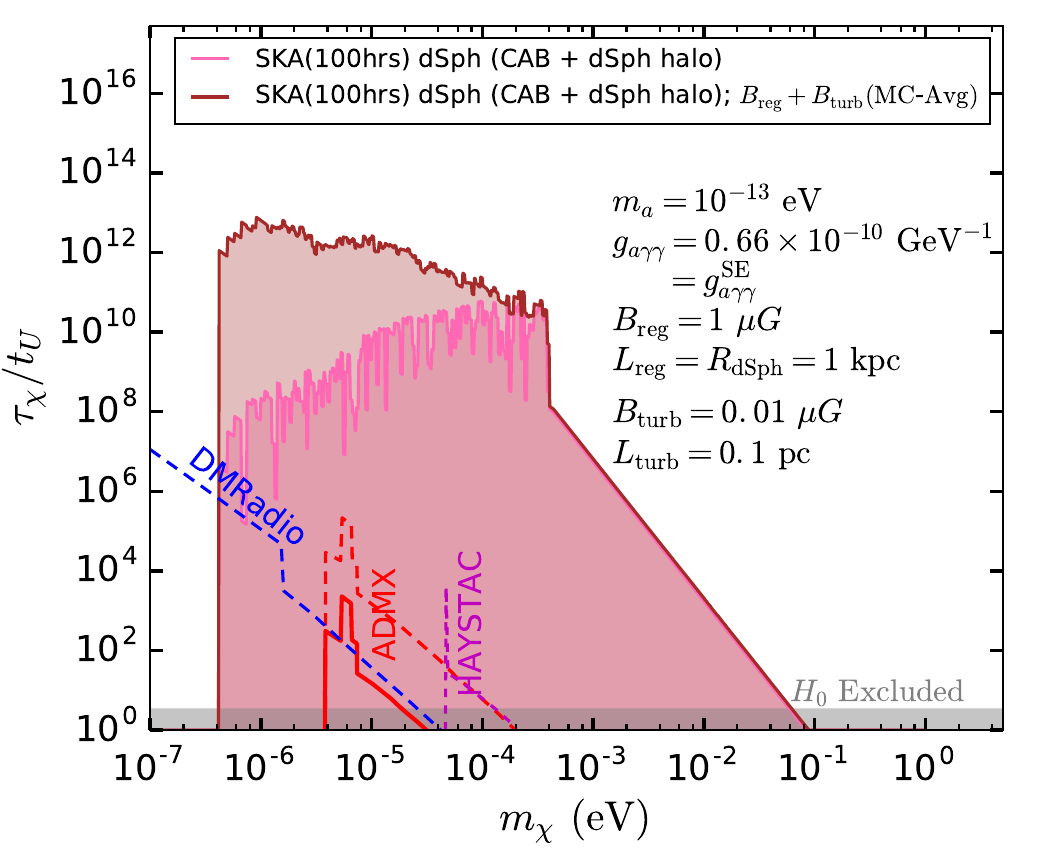}
\includegraphics[width=6cm]{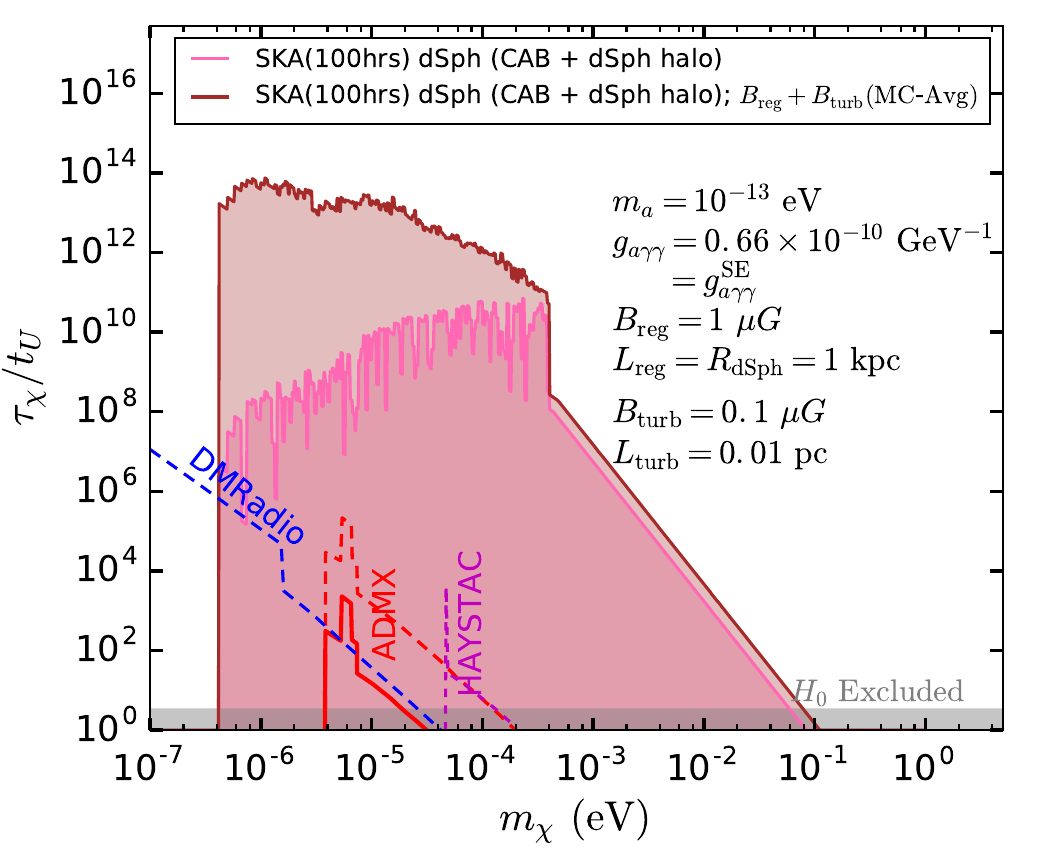}
\caption{The projected SKA reach in the $m_\chi-\tau_\chi$ plane  after 100 hours of Seg I observations with the turbulent magnetic field parameters and color coding as in Fig. \ref{fig:limit_g_ma_Ptot}, while the axion and the regular magnetic field parameters are as in Fig.~\ref{fig:flux_Ptot}. 
The color coding for the other exclusions is as in Fig. \ref{fig:limit_tau_mdm}. 
}
\label{fig:limit_tau_mdm_Ptot}
\end{figure*}

The effect of the turbulent magnetic field component on the SKA reach  is clearly visible in Figs.~\ref{fig:limit_g_ma_Ptot}  and \ref{fig:limit_tau_mdm_Ptot}. These show the projected SKA reach (brown) in the $g_{a \gamma \gamma} - m_a$ and $m_\chi-\tau_\chi$ planes, respectively, after 100 hours of Seg I  observation, for the three representative turbulent magnetic field parameter sets  in Eq. \eqref{eq:Bturb}, to be compared with the corresponding SKA reach assuming just the regular magnetic field (pink). 
All the parameters that are not varied, including the regular magnetic field parameters, are  as in Fig.~\ref{fig:flux_Ptot}. 
The presence of a turbulent magnetic field component  improves significantly the projected SKA sensitivities on $g_{a\gamma\gamma}$, especially for axion masses above $10^{-13}$ eV, see Fig.~\ref{fig:limit_g_ma_Ptot}.  Similarly, if there is a turbulent magnetic field component in Seg I, this will increase the SKA sensitivity to DM decay time,  especially for 
$m_{\chi}$ in the range $4 \times 10^{-7} \sim 2 \times 10^{-4}$ eV, see Fig. \ref{fig:limit_tau_mdm_Ptot}.

\subsection{Projected SKA reach including stimulated emission}
\label{sec:proj:stim}
\begin{figure}[t]
\centering 
\includegraphics[width=7cm]{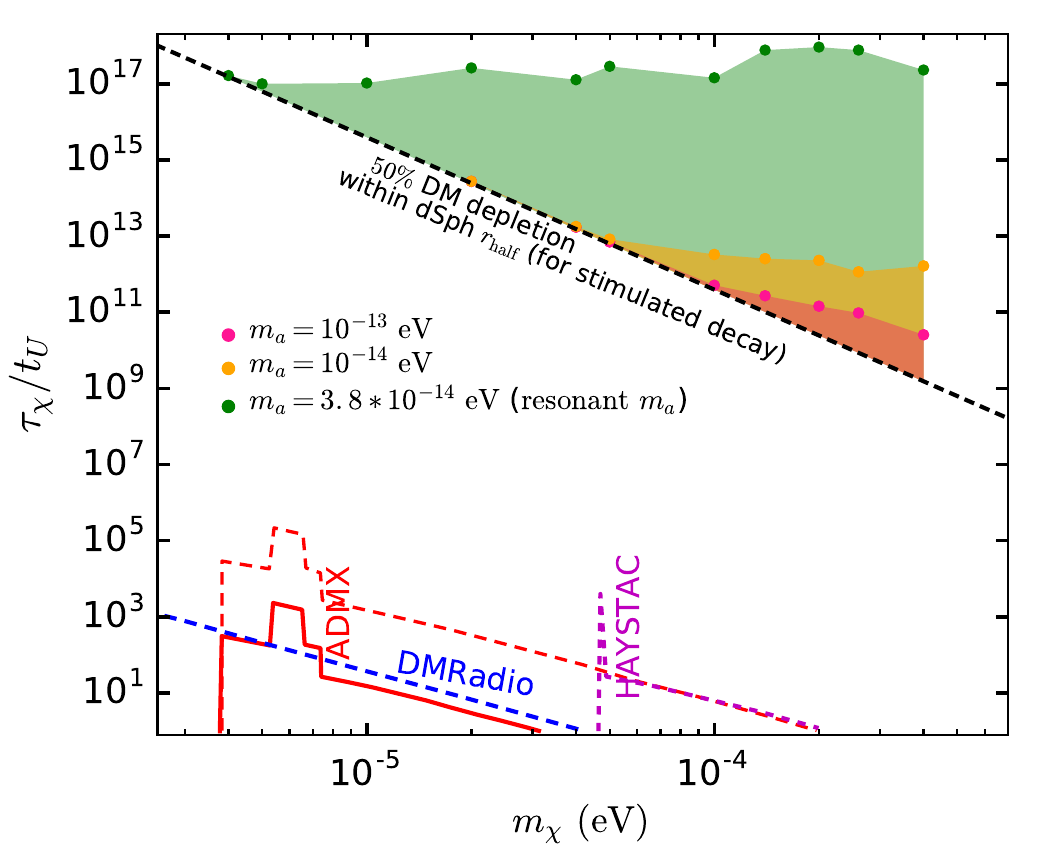}
\includegraphics[width=7cm]{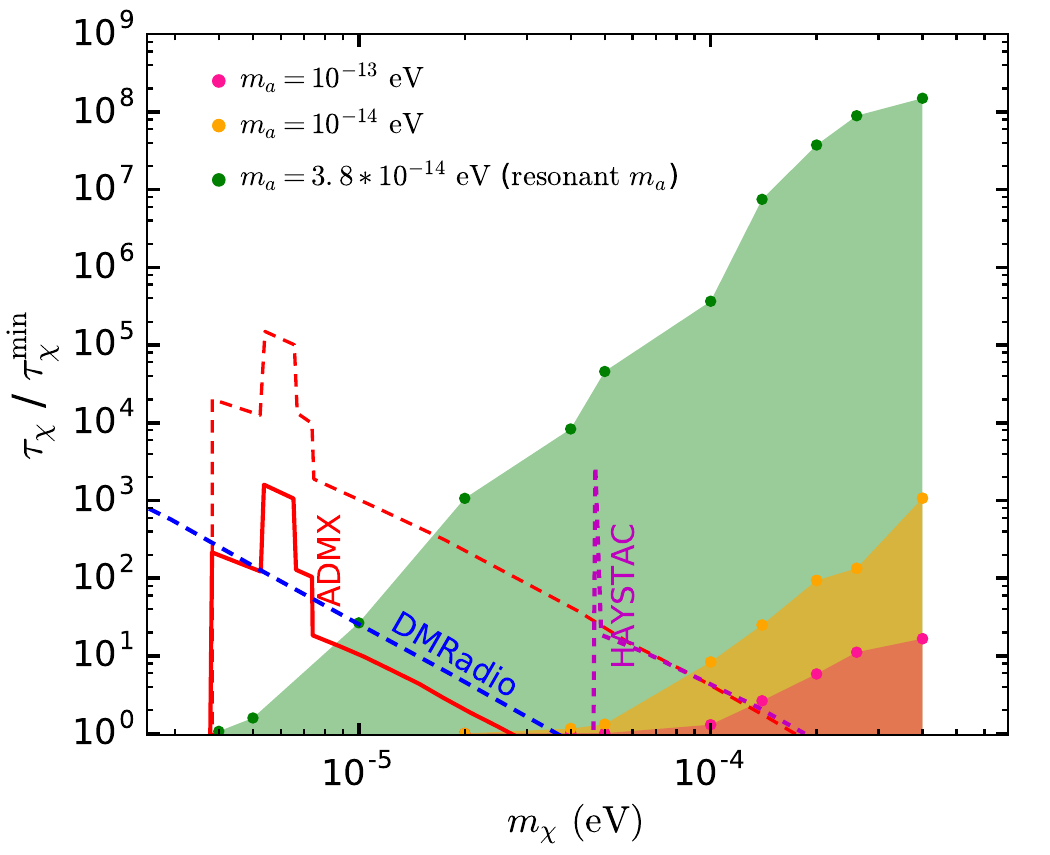}
\caption{The 100h SKA reach in $\tau_\chi/t_U$ (left) and $\tau_\chi/\tau_\chi^{\rm min}$ (right) for three axion masses, $m_a = \{1,3.8,10\} \times 10^{-14}$\,eV (\{orange, green, pink\}), using only the dominant contribution to the radio flux density, i.e., from DM decays in dSph. The other axion and astrophysical parameters, as well as the (projected) exclusions from other experiments, are the same as those used in the left panel of Fig.~\ref{fig:limit_tau_mdm}. The parameter region above the dashed black line in the left panel corresponds to slow enough DM decays, see discussion in the main text. 
}
\label{fig:tau0_t0_mx_ska}
\end{figure}

Finally, let us turn to the predictions of SKA reach that includes the effect of stimulated emissions. The 
radio flux density due to converted axions is then given by Eq. \eqref{eq:radio_flux_dsph}, after replacing the DM decay time in the expression for $\left.{dn_a}/{dt d\omega}\right\vert_{\rm dSph}$, Eq.~\eqref{eq:dna_dw_dsph}, with the effective decay time, $\tau_\chi\to 1/\Gamma_{\rm eff}$. 
The radio flux density is then given by 
\begin{equation}
S(\nu) = P_{a\gamma}(\omega) \hspace{1mm} 
\times \frac{1}{4 \pi \nu} \int_{\Omega} \int_{\rm l.o.s} d\Omega \hspace{1mm} ds \hspace{1mm} \omega^2 \Gamma_{\rm eff}(s) \frac{dN_a}{d\omega} n_{\chi} 
\hspace{2mm}
\text{(in the units of $\rm erg\; cm^{-2} s^{-1} Hz^{-1}$)},
\label{eq:stimulated_radio_flux_dsph}
\end{equation}
where the energy spread  in the photon signal, encoded in $dN_a/d\omega$, is due to both the dSph DM velocity spread  as well as due to the velocity of the Earth, and is modeled via Gaussian smearing as in Section~\ref{sec:dSph_contribution}. Here $\nu = \omega / 2\pi$ is the frequency of the detected photon. For simplicity, we only consider the case of a homogeneous magnetic field in the dSph, so that the axion$\rightarrow$photon conversion probability $P_{a\gamma}(\omega)$ is given by the expressions in Section~\ref{sec:single:domain}, taking dSph as a single magnetic domain (cf. also the analysis in Section \ref{sec:coh:magn:field}). 
Just like before (cf. also the analysis in Section \ref{sec:radio_flux}), in the numerical estimates of the flux density $S$ we use the conventional units of Jy, 
where 1 Jy = $10^{-23}$ $\rm erg$ $\rm cm^{-2} s^{-1} Hz^{-1}$. 

Note that the effective decay width in \eqref{eq:stimulated_radio_flux_dsph} is position dependent, since it depends on the local axion density, $n_a$, see \eqref{eq:Gamma_eff}. It is calculated in the limit of slowly varying axion density using the iterative procedure detailed in Section \ref{sec:stimulated_emission} and in Appendix \ref{sec:app:stimulated}.  Using this procedure also outside its region of validity one finds a runaway behavior. If this is a physical effect most of the axion and DM parameter region that can be probed by SKA (and other experiments) is already excluded, since it results in DM decaying too quickly. This can be seen from Fig. \ref{fig:tau0_t0_mx_ska} (left) where we show the 100h SKA reach in the $m_\chi-\tau_\chi$ plane for three axion masses: $m_a = 3.8 \times 10^{-14}$\,eV (green) that leads to resonant conversion and thus the highest SKA reach, and two axion masses, below and above the resonant conversion, $m_a = 10^{-14}$\,eV (pink)  and $m_a = 10^{-13}$\,eV (orange), respectively. Other axion and astrophysical parameters are the same as those used in the left panel of Fig.~\ref{fig:limit_tau_mdm}. The same is true for the plotted exclusion lines from the other experiments, i.e., they are as in  Fig.~\ref{fig:limit_tau_mdm} and in particular ignore the effects of stimulated emissions.  In calculating the SKA reach we 
considered for numerical expedience only the contribution from the decay of 
local DM particles in the dSph halo, i.e., we neglected the CAB contribution, which explains the sharp cut-off of SKA sensitivity at high DM masses (cf. Fig. \ref{fig:limit_tau_mdm}). In fact, since estimating $\Gamma_{\rm eff}$ is rather numerically intensive, the SKA reach calculation shown in Fig. \ref{fig:tau0_t0_mx_ska} is rather coarse grained, with dots denoting the calculated values, and the remainder obtained through linear interpolation to guide the eye. 

The dashed black line in Fig.~\ref{fig:tau0_t0_mx_ska} (left) denotes the maximum allowed value of $\Gamma_{\chi}$ (minimal allowed value of $\tau_\chi$), which we define to  be the $\Gamma_\chi$ for which 50\% of DM contained within half-light radius of the dSph decays during the evolution of dSph until the present time 
(this corresponds to roughly $\sim$10\% depletion of total dSph DM mass). From numerical analysis we obtain,
\begin{equation}
   \Gamma_{\chi}^{\rm max} (m_{\chi})\equiv1/\tau_\chi^{\rm min}  \simeq 5.86 \times 10^{-38} \times 
    \left( \frac{m_{\chi}}{1\;\mu\rm eV} \right)^4 \:\: s^{-1} .
\label{eq:G_max_defn}
\end{equation}
Note that while this line is in the  region outside the validity of our iterative procedure, it is not very far away from it (cf. Fig. \ref{fig:fahat}).
For reader's convenience we also show the projected 100h SKA reach in terms of normalized DM decay time, $\tau_\chi/\tau_\chi^{\rm min}$, in Fig. \ref{fig:tau0_t0_mx_ska} (right). We see that a discovery of axion by the SKA would require a coincidence, a DM that has just the right decay time that it does not lead to too quick DM decays, but still small enough that it gives an observable radio signal. The other option is that the axion mass is such that resonant conversion is possible. Two comments are in order, though. First of all, as already discussed in Section \ref{sec:stimulated_emission}, establishing whether shorter DM decay times really do lead to a physical runaway behavior would require a more detailed numerical simulation that takes into account the possibility of a rapidly changing axion field. Secondly, even if the runaway behavior  is physical, improving the SKA reach (either by an increased exposure time or relying on spectral features) would drive the observations away from the ``just-now''  parameter region.

\section{Summary and conclusions}
\label{sec:conclusion}

In this manuscript we explored a rather simple modification of the minimal DM scenario, allowing for DM to be accompanied by a light pseudoscalar, an axion $a$, such that DM is unstable and decays through the $\chi \to a a$ channel (for simplicity we assume DM to be a scalar). This is still a rather generic possibility given that a dark sector with a spontaneously broken global symmetry will result in a pNGB, i.e., such a light pseudoscalar.

 We were most interested in the possibility that axion couples to photons, in which case it can be searched for in terrestrial experiments. DM decays produce relativistic axions, which then convert to photons  in the 
magnetic field of a galaxy such as a dSph. As we showed in the paper, this process 
can
 lead to a detectable signal in the upcoming SKA radio telescope for DM masses in the range few$\times10^{-7}\,\text{eV}<m_{\chi}<
\text{few}\times10^{-2}$\,eV.  Assuming a 100\,h SKA observation of a local dSph Seg I the projected  sensitivity to axion-photon coupling $g_{a\gamma\gamma}$   is, for axion masses below $m_a \lesssim 10^{-13}$\,eV, expected to be orders of magnitudes better than for dedicated 
current and planned axion dark matter experiments, such as ADMX, HAYSTAC, etc. These parameter range is also allowed by the bounds obtained from  
CAST helioscope, SN 1987A constraints, and the  CMB measurements. For higher values of $m_a$ the SKA 
sensitivity weakens, because the axion-photon conversion 
probability in the regular magnetic field of a galaxy drops rapidly, if 
 $m_a$ is increased. 

There are a number of uncertainties that enter the SKA projections, and whose effect we have estimated in the paper. The first issue is the dependence of radio flux on the value of the magnetic field in the dSph. The most conservative assumption is that the regular magnetic field is small, an order of magnitude smaller than expectations, and has negligible turbulent component, giving projections in Fig. \ref{fig:limit_g_ma} (right) and \ref{fig:limit_tau_mdm} (right).  A turbulent magnetic field component, however, is expected to be large enough in dSph Seg I, such that the enhancement of the radio flux is expected, leading to results in Section \ref{sec:turbulent_conversion}. Finally, the axion density in the dSph is large enough that it can lead to significant Bose enhancement of the DM decay rates. To estimate the effect of such stimulated emission we used an iterative procedure that applies in the regime of a slowly varying axion density, in which we obtained Bose enhancements of the DM decay rates of up to ${\mathcal O}(10^3)$. Unfortunately, most of the DM and axion parameter regime potentially relevant for SKA lies outside the validity of this approximation, and thus further work is needed. It is possible that the stimulated emission results in DM decays that are exponentially enhanced, such that in large regions of parameter space DM is unstable on cosmological timescales. Such runaway behavior is for instance obtained, if we use our iterative method outside its regime of validity. It is, however, also possible that a more fine grained time dependent simulation would result in quasistable solutions, similar to what we found with our iterative method when restricted to its regime of validity. While the issue of the effect of stimulated emission is not settled, it is useful to note that SKA can still probe regions of axion parameters for which our simulations are reliable, see Fig. \ref{fig:tau0_t0_mx_ska}, and further discussion in Section \ref{sec:proj:stim}.

To conclude, searches for DM induced relativistic axion signals 
using radio telescopes can play an important role in axion physics, and can be complementary to other probes such as the axion haloscopes and helioscopes.

\section*{Acknowledgements}
We thank Nick Rodd for useful discussions. 
The research of AK was supported by the National Research 
Foundation of Korea (NRF) funded by the Ministry of Education 
through the Center for Quantum Space Time (CQUeST) of Sogang University 
with Grant No. 2020R1A6A1A03047877 and by the Ministry of Science and ICT with grant
number 2021R1F1A1057119.	
TK acknowledges support in the form of Junior Research Fellowship from the Council of Scientific \& Industrial Research (CSIR), Government of India. 
SR acknowledges support from the APS-IUSSTF Professorship Award and the Department of Physics, University of Cincinnati for facilitating his visit to University of Cincinnati where this work was initiated. 
JZ acknowledges support in part by the DOE grant de-sc0011784 and NSF OAC-2103889. 
AK acknowledges the stay at School of Physical Sciences, Indian Association for the Cultivation of Science (IACS) where the initial part of this project was formulated.

\begin{appendices}

\section{Iterative solutions for axion occupation number}
\label{sec:app:stimulated}
In this appendix we give further details on finding the axion occupation number $\hat f_a(r)$ in \eqref{eq:fa} by iteratively solving for $n_a(r)$ in~\eqref{eq:na_r}. We use two procedures, the {\em default} one, which was then used in the numerical analysis in the main text, and the {\em alternative} one, which is used as a consistency check. Both rely on the assumption that axion density is changing only slowly. 

\paragraph{The default procedure.} Since the change in $n_a$ is slow compared to the time it takes for axions to traverse the dSph, we can solve Eq.~\eqref{eq:na_r} iteratively to find a quasi-equilibrium value of $n_a(r)$ at present. 
As a time step we use $\Delta t_{\rm dSph}=R_{\rm dSph}/c$, and assume that at $t=0$ the initial axion number density in the dSph is zero, $n_a(t=0)=0$. The axion number density at time $t=(i+1)\Delta t_{\rm dSph}$ is then given by the l.h.s. of Eq.~\eqref{eq:na_r}, where on the r.h.s. the effective decay width is calculated with $n_a$ from the $i$-th time step, $\Gamma_{{\rm eff}, i} = \Gamma_{\rm eff}(n_{a,i})$,
 \begin{equation}
        n_{a,i+1} (r) = \int d\Omega \int_{\rm l.o.s} ds \frac{\Gamma_{{\rm eff},i} N_a n_{\chi,i}}{4 \pi}.
\label{eq:na_i}
\end{equation}
In the calculation we also reduce appropriately the DM number density,
\begin{equation}
    n_{\chi,i+1} = n_{\chi,i} \exp[-\Gamma_{{\rm eff},i}\Delta t_{\rm dSph}].
\label{eq:nx_i}
\end{equation}

\begin{figure}[t]
    \centering
    \includegraphics[width=7cm]{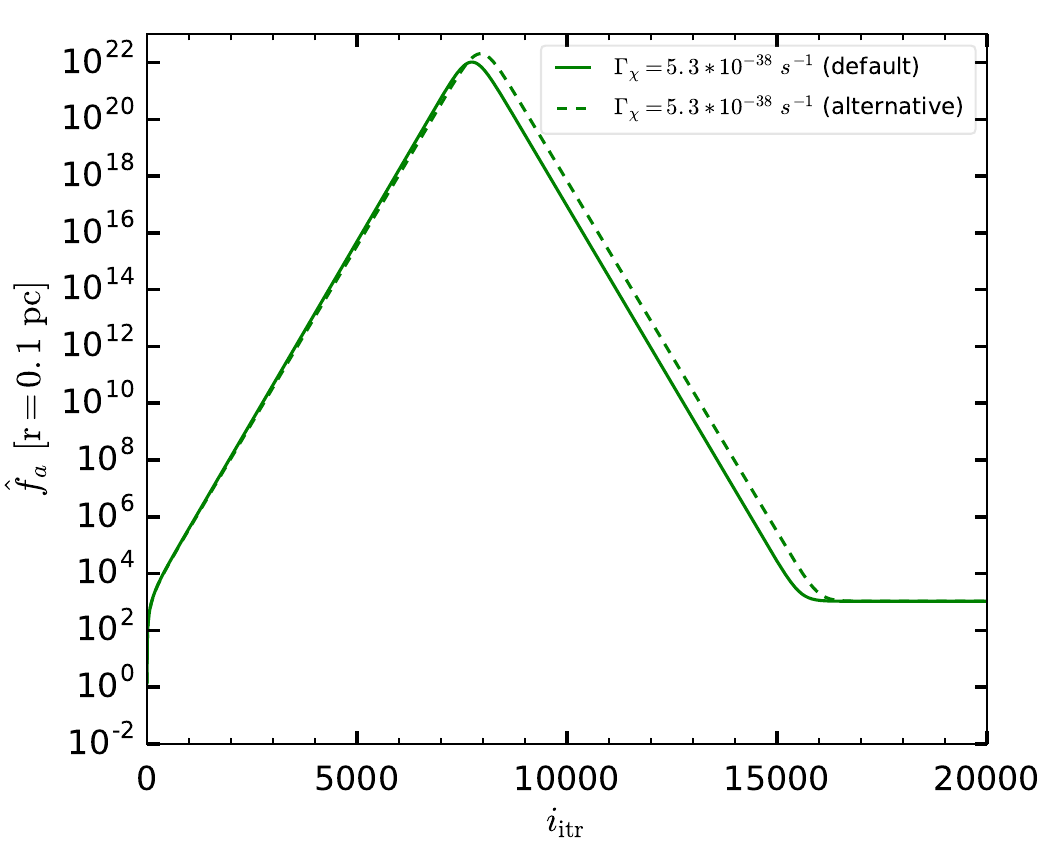}
    \includegraphics[width=7cm]{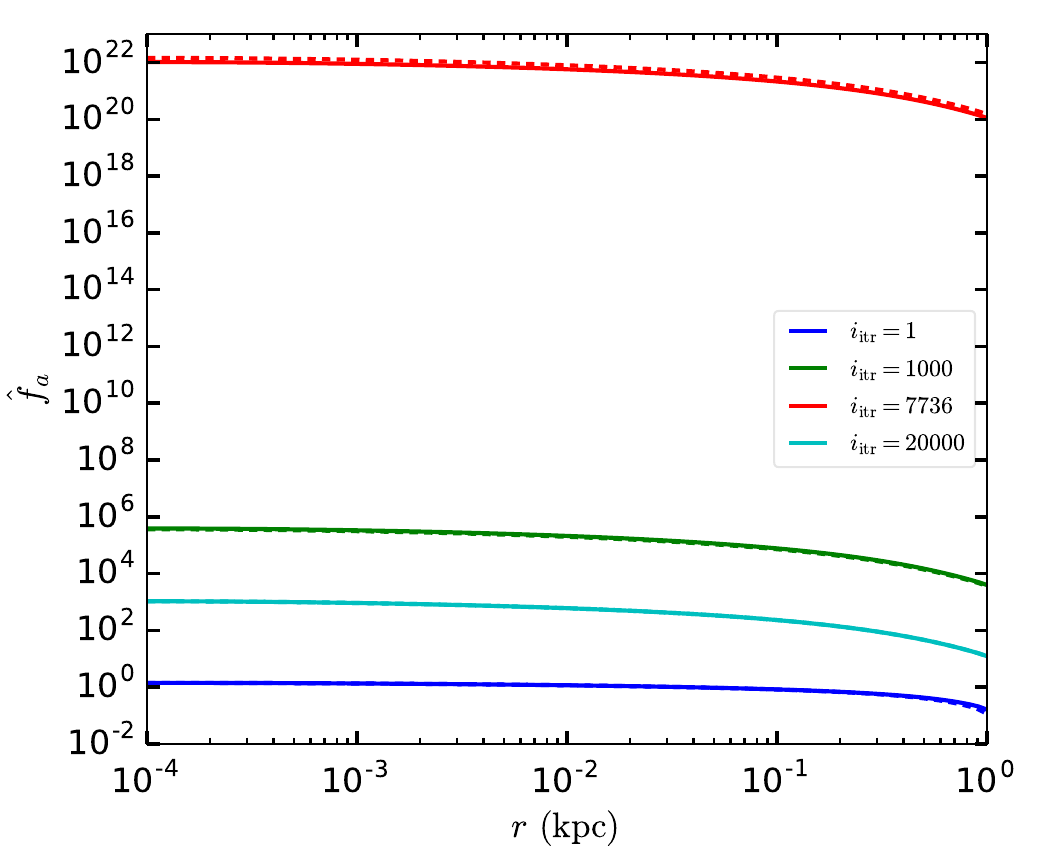}
     \includegraphics[width=7cm]{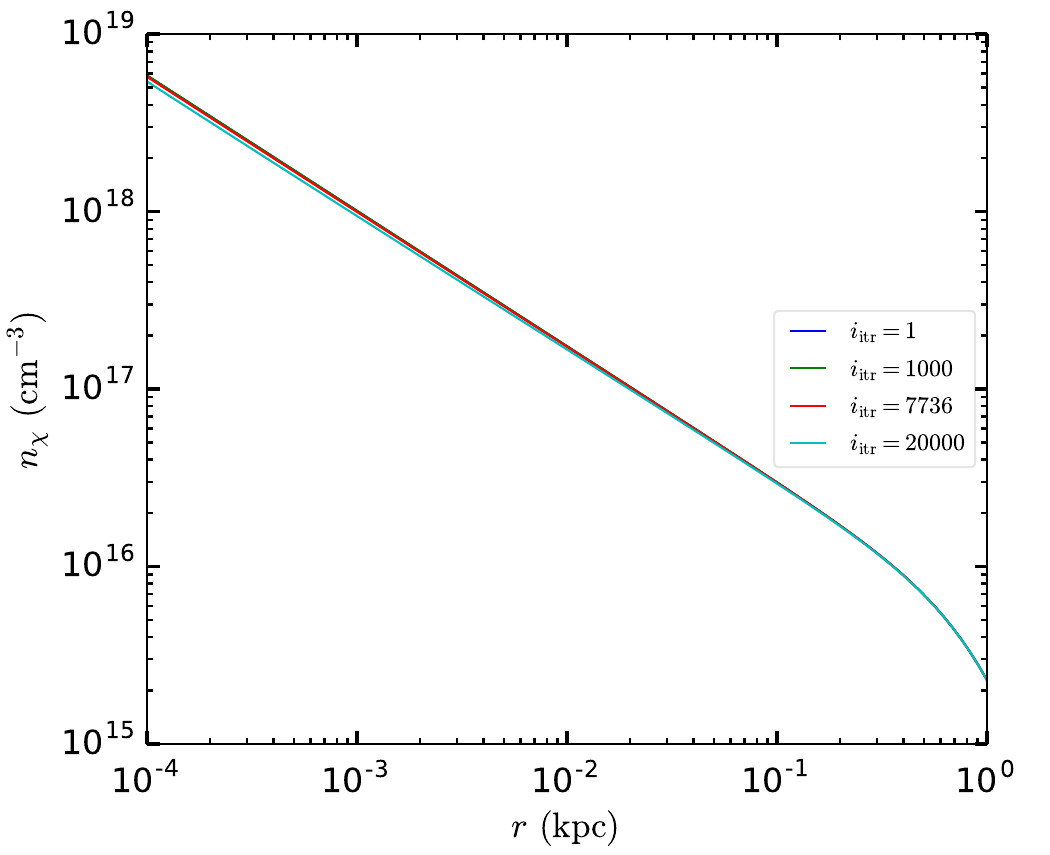}
    \caption{
    {\bf Left top:} The variation of axion occupation number $\hat f_a (r = 0.1 \; \rm pc)$ with iteration step, $i_{\rm itr}$, in the default (solid green line) and alternative (dashed green line) methods, for $m_{\chi} = 1$ $\mu$eV and $\Gamma_\chi=5.3\cdot 10^{-38}\,\text{s}^{-1}$.  {\bf Right top:} Radial profile of the axion occupation number $\hat f_a$ inside the dSph for four different iteration steps $i_{\rm itr}$ (step $i_{\rm itr} = 7736$ represents the peak value for most $r$). {\bf Bottom:} The corresponding changes in DM number density, $n_\chi$. 
    }
    \label{fig:na_new_old_5.3}
\end{figure}

The results of this procedure for $m_{\chi} = 1$ $\mu$eV and $\Gamma_\chi=5.3\cdot 10^{-38}\,\text{s}^{-1}$  are shown in Fig.~\ref{fig:na_new_old_5.3}, top left panel (green solid line). In this example, after about $15\cdot 10^3$ iterations we reach the equilibrium value of $n_a$ that corresponds to $\hat f_a\sim {\mathcal O}(10^3)$ in the inner part of dSph and $\hat f_a\sim {\mathcal O}(10)$ in the outskirts. The equilibrium value changes slowly with time since DM slowly decays, which results in decreasing equilibrium value of $n_a$. This change is very slow, however, since the effective DM decay time is still very long,  $1/\Gamma_{\rm eff}\gg t_U$. Note also, that the iteration step at which the equilibrium is reached, $i_{\rm itr}\simeq 15\cdot 10^3$, corresponds to a time of about $5\cdot 10^7\,$years, which is much less than the age of the dSph galaxy. The calculation of the equilibrium value of $\hat f_a$ also agrees with the result obtained from the alternative iterative procedure (dashed green line in Fig. \ref{fig:na_new_old_5.3}, top left panel), which we describe next. 

\paragraph{The alternative procedure.} 

To test that the assumption of a slowly varying axion number density is valid for a particular parameter region, we split the dSph into two halves, and use $\hat f_a$ values from two different iterative steps, when calculating $\Gamma_{\rm eff}$. That is, as in the default procedure we set in the $0$-th step  $n_{a,0} = 0$ and set the initial DM density to $n_{\chi,0}=n_{\rm DM}$. To calculate the axion density at a given point \textbf{p}, we then draw lines of sights from \textbf{p} into all directions and divide the galaxy into two zones: (i) the  near zone: $s < R_{\rm dSph}$, and (ii) the far zone: $s > R_{\rm dSph}$. Note that this division of dSph is different for different positions of \textbf{p}. The $n_a$ in the $(i+1)$-th step is then calculated using 
    \begin{equation}
        n_{a,i+1} (r) = \int d\Omega \int_{\rm l.o.s,\;near} ds \frac{\Gamma_{{\rm eff},i} N_a n_{\chi,i}}{4 \pi} + \int d\Omega \int_{\rm l.o.s,\;far} ds \frac{\Gamma_{{\rm eff},i-1} N_a n_{\chi,i-1}}{4 \pi},
    \end{equation}
   while  the change in DM density is given by 
    \begin{equation}
       \delta n_{\chi,i+1} = - n_{\chi,i} \delta t \Gamma_{{\rm eff}, i}.
    \end{equation}
The time step is taken to be $\delta t = \Delta t_{\rm dSph}$ and $\Gamma_{{\rm eff}, i} = \Gamma_{\rm eff}(n_{a,i})$. 
    The motivation for using the split of dSph into near and far zones is that it takes time for axions to reach the point \textbf{p}, and we thus use retarded values of $n_a$ in the far zone. If the $n_a$ changes slowly, the effect of retardation is negligible and one would obtain the same result for equilibrium value of $n_a$ as in the default procedure, where the change in $n_a$ for the whole dSph is treated as being instantaneous. In view of this interpretation we thus take $n_a$ in the first step to be given by, 
\begin{equation}
        n_{a,1} = \int d\Omega \int_{\rm l.o.s,\;near} ds \frac{\Gamma_{\rm eff,0}N_an_{\chi,0}}{4 \pi},
    \end{equation}
    that is, we assume that the axions from the far zone have not yet reached point \textbf{p}. 
    
    The results for $\hat f_a$ obtained from the alternative procedure for the  $m_{\chi} = 1$ $\mu$eV, $\Gamma_\chi=5.3\cdot 10^{-38}\,\text{s}^{-1}$ example are shown in Fig. \ref{fig:na_new_old_5.3}, top left panel (green dashed line). Since in this case the change in $n_a$ is slower than the propagation time in dSph the equilibrium values of $n_a$ obtained with the default and alternative procedures agree.

\end{appendices}

\bibliographystyle{JHEP}
\bibliography{biblio}

\providecommand{\href}[2]{#2}\begingroup\raggedright\begin{thebibliography}{10}

\bibitem{Kim:2008hd}
J.~E. Kim and G.~Carosi, \emph{{Axions and the Strong CP Problem}},
  \href{https://doi.org/10.1103/RevModPhys.82.557}{\emph{Rev. Mod. Phys.}
  {\bfseries 82} (2010) 557--602},
  [\href{https://arxiv.org/abs/0807.3125}{{\ttfamily 0807.3125}}].

\bibitem{Marsh:2015xka}
D.~J.~E. Marsh, \emph{{Axion Cosmology}},
  \href{https://doi.org/10.1016/j.physrep.2016.06.005}{\emph{Phys. Rept.}
  {\bfseries 643} (2016) 1--79},
  [\href{https://arxiv.org/abs/1510.07633}{{\ttfamily 1510.07633}}].

\bibitem{DiLuzio:2020wdo}
L.~Di~Luzio, M.~Giannotti, E.~Nardi and L.~Visinelli, \emph{{The landscape of
  QCD axion models}},
  \href{https://doi.org/10.1016/j.physrep.2020.06.002}{\emph{Phys. Rept.}
  {\bfseries 870} (2020) 1--117},
  [\href{https://arxiv.org/abs/2003.01100}{{\ttfamily 2003.01100}}].

\bibitem{1992SvJNP..55.1063B}
Z.~G. {Berezhiani}, A.~S. {Sakharov} and M.~Y. {Khlopov}, \emph{{Primordial
  background of cosmological axions.}}, {\emph{Soviet Journal of Nuclear
  Physics} {\bfseries 55} (July, 1992) 1063--1071}.

\bibitem{Preskill:1982cy}
J.~Preskill, M.~B. Wise and F.~Wilczek, \emph{{Cosmology of the Invisible
  Axion}}, \href{https://doi.org/10.1016/0370-2693(83)90637-8}{\emph{Phys.
  Lett. B} {\bfseries 120} (1983) 127--132}.

\bibitem{Abbott:1982af}
L.~F. Abbott and P.~Sikivie, \emph{{A Cosmological Bound on the Invisible
  Axion}}, \href{https://doi.org/10.1016/0370-2693(83)90638-X}{\emph{Phys.
  Lett. B} {\bfseries 120} (1983) 133--136}.

\bibitem{Dine:1982ah}
M.~Dine and W.~Fischler, \emph{{The Not So Harmless Axion}},
  \href{https://doi.org/10.1016/0370-2693(83)90639-1}{\emph{Phys. Lett. B}
  {\bfseries 120} (1983) 137--141}.

\bibitem{Conlon:2013isa}
J.~P. Conlon and M.~C.~D. Marsh, \emph{{The Cosmophenomenology of Axionic Dark
  Radiation}}, \href{https://doi.org/10.1007/JHEP10(2013)214}{\emph{JHEP}
  {\bfseries 10} (2013) 214},
  [\href{https://arxiv.org/abs/1304.1804}{{\ttfamily 1304.1804}}].

\bibitem{Marsh:2013opc}
M.~C.~D. Marsh, \emph{{Hints of a Cosmic Axion Background}},  in \emph{{9th
  Patras Workshop on Axions, WIMPs and WISPs}}, pp.~159--163, 2013,
  \href{https://doi.org/10.3204/DESY-PROC-2013-04/marsh_david}{DOI}.

\bibitem{Dror:2021nyr}
J.~A. Dror, H.~Murayama and N.~L. Rodd, \emph{{Cosmic axion background}},
  \href{https://doi.org/10.1103/PhysRevD.103.115004}{\emph{Phys. Rev. D}
  {\bfseries 103} (2021) 115004},
  [\href{https://arxiv.org/abs/2101.09287}{{\ttfamily 2101.09287}}].

\bibitem{Langhoff:2022bij}
K.~Langhoff, N.~J. Outmezguine and N.~L. Rodd, \emph{{The Irreducible Axion
  Background}},  \href{https://arxiv.org/abs/2209.06216}{{\ttfamily
  2209.06216}}.

\bibitem{Ouellet:2018beu}
J.~L. Ouellet et~al., \emph{{First Results from ABRACADABRA-10 cm: A Search for
  Sub-$\mu$eV Axion Dark Matter}},
  \href{https://doi.org/10.1103/PhysRevLett.122.121802}{\emph{Phys. Rev. Lett.}
  {\bfseries 122} (2019) 121802},
  [\href{https://arxiv.org/abs/1810.12257}{{\ttfamily 1810.12257}}].

\bibitem{Ouellet:2019tlz}
J.~L. Ouellet et~al., \emph{{Design and implementation of the ABRACADABRA-10 cm
  axion dark matter search}},
  \href{https://doi.org/10.1103/PhysRevD.99.052012}{\emph{Phys. Rev. D}
  {\bfseries 99} (2019) 052012},
  [\href{https://arxiv.org/abs/1901.10652}{{\ttfamily 1901.10652}}].

\bibitem{Silva-Feaver:2016qhh}
M.~Silva-Feaver et~al., \emph{{Design Overview of DM Radio Pathfinder
  Experiment}}, \href{https://doi.org/10.1109/TASC.2016.2631425}{\emph{IEEE
  Trans. Appl. Supercond.} {\bfseries 27} (2017) 1400204},
  [\href{https://arxiv.org/abs/1610.09344}{{\ttfamily 1610.09344}}].

\bibitem{DMRadio:2022pkf}
{\scshape DMRadio} collaboration, L.~Brouwer et~al., \emph{{DMRadio-m$^3$: A
  Search for the QCD Axion Below $1\,\mu$eV}},
  \href{https://arxiv.org/abs/2204.13781}{{\ttfamily 2204.13781}}.

\bibitem{Brouwer:2022bwo}
{\scshape DMRadio} collaboration, L.~Brouwer et~al., \emph{{Proposal for a
  definitive search for GUT-scale QCD axions}},
  \href{https://doi.org/10.1103/PhysRevD.106.112003}{\emph{Phys. Rev. D}
  {\bfseries 106} (2022) 112003},
  [\href{https://arxiv.org/abs/2203.11246}{{\ttfamily 2203.11246}}].

\bibitem{ADMX:2003rdr}
{\scshape ADMX} collaboration, S.~J. Asztalos et~al., \emph{{An Improved RF
  cavity search for halo axions}},
  \href{https://doi.org/10.1103/PhysRevD.69.011101}{\emph{Phys. Rev. D}
  {\bfseries 69} (2004) 011101},
  [\href{https://arxiv.org/abs/astro-ph/0310042}{{\ttfamily
  astro-ph/0310042}}].

\bibitem{ADMX:2018gho}
{\scshape ADMX} collaboration, N.~Du et~al., \emph{{A Search for Invisible
  Axion Dark Matter with the Axion Dark Matter Experiment}},
  \href{https://doi.org/10.1103/PhysRevLett.120.151301}{\emph{Phys. Rev. Lett.}
  {\bfseries 120} (2018) 151301},
  [\href{https://arxiv.org/abs/1804.05750}{{\ttfamily 1804.05750}}].

\bibitem{ADMX:2019uok}
{\scshape ADMX} collaboration, T.~Braine et~al., \emph{{Extended Search for the
  Invisible Axion with the Axion Dark Matter Experiment}},
  \href{https://doi.org/10.1103/PhysRevLett.124.101303}{\emph{Phys. Rev. Lett.}
  {\bfseries 124} (2020) 101303},
  [\href{https://arxiv.org/abs/1910.08638}{{\ttfamily 1910.08638}}].

\bibitem{PhysRevD.96.123008}
B.~M. Brubaker, L.~Zhong, S.~K. Lamoreaux, K.~W. Lehnert and K.~A. van Bibber,
  \emph{Haystac axion search analysis procedure},
  \href{https://doi.org/10.1103/PhysRevD.96.123008}{\emph{Phys. Rev. D}
  {\bfseries 96} (Dec, 2017) 123008}.

\bibitem{HAYSTAC:2018rwy}
{\scshape HAYSTAC} collaboration, L.~Zhong et~al., \emph{{Results from phase 1
  of the HAYSTAC microwave cavity axion experiment}},
  \href{https://doi.org/10.1103/PhysRevD.97.092001}{\emph{Phys. Rev. D}
  {\bfseries 97} (2018) 092001},
  [\href{https://arxiv.org/abs/1803.03690}{{\ttfamily 1803.03690}}].

\bibitem{Adair:2022rtw}
C.~M. Adair et~al., \emph{{Search for Dark Matter Axions with CAST-CAPP}},
  \href{https://doi.org/10.1038/s41467-022-33913-6}{\emph{Nature Commun.}
  {\bfseries 13} (2022) 6180},
  [\href{https://arxiv.org/abs/2211.02902}{{\ttfamily 2211.02902}}].

\bibitem{Caldwell:2016dcw}
{\scshape MADMAX Working Group} collaboration, A.~Caldwell, G.~Dvali,
  B.~Majorovits, A.~Millar, G.~Raffelt, J.~Redondo et~al., \emph{{Dielectric
  Haloscopes: A New Way to Detect Axion Dark Matter}},
  \href{https://doi.org/10.1103/PhysRevLett.118.091801}{\emph{Phys. Rev. Lett.}
  {\bfseries 118} (2017) 091801},
  [\href{https://arxiv.org/abs/1611.05865}{{\ttfamily 1611.05865}}].

\bibitem{Millar:2016cjp}
A.~J. Millar, G.~G. Raffelt, J.~Redondo and F.~D. Steffen, \emph{{Dielectric
  Haloscopes to Search for Axion Dark Matter: Theoretical Foundations}},
  \href{https://doi.org/10.1088/1475-7516/2017/01/061}{\emph{JCAP} {\bfseries
  01} (2017) 061}, [\href{https://arxiv.org/abs/1612.07057}{{\ttfamily
  1612.07057}}].

\bibitem{Planck:2018vyg}
{\scshape Planck} collaboration, N.~Aghanim et~al., \emph{{Planck 2018 results.
  VI. Cosmological parameters}},
  \href{https://doi.org/10.1051/0004-6361/201833910}{\emph{Astron. Astrophys.}
  {\bfseries 641} (2020) A6},
  [\href{https://arxiv.org/abs/1807.06209}{{\ttfamily 1807.06209}}].

\bibitem{Baumann:2016wac}
D.~Baumann, D.~Green and B.~Wallisch, \emph{{New Target for Cosmic Axion
  Searches}}, \href{https://doi.org/10.1103/PhysRevLett.117.171301}{\emph{Phys.
  Rev. Lett.} {\bfseries 117} (2016) 171301},
  [\href{https://arxiv.org/abs/1604.08614}{{\ttfamily 1604.08614}}].

\bibitem{Eby:2021ece}
J.~Eby, S.~Shirai, Y.~V. Stadnik and V.~Takhistov, \emph{{Probing relativistic
  axions from transient astrophysical sources}},
  \href{https://doi.org/10.1016/j.physletb.2021.136858}{\emph{Phys. Lett. B}
  {\bfseries 825} (2022) 136858},
  [\href{https://arxiv.org/abs/2106.14893}{{\ttfamily 2106.14893}}].

\bibitem{Conlon:2013txa}
J.~P. Conlon and M.~C.~D. Marsh, \emph{{Excess Astrophysical Photons from a
  0.1\textendash{}1 keV Cosmic Axion Background}},
  \href{https://doi.org/10.1103/PhysRevLett.111.151301}{\emph{Phys. Rev. Lett.}
  {\bfseries 111} (2013) 151301},
  [\href{https://arxiv.org/abs/1305.3603}{{\ttfamily 1305.3603}}].

\bibitem{Moroi:2018vci}
T.~Moroi, K.~Nakayama and Y.~Tang, \emph{{Axion-photon conversion and effects
  on 21 cm observation}},
  \href{https://doi.org/10.1016/j.physletb.2018.07.002}{\emph{Phys. Lett. B}
  {\bfseries 783} (2018) 301--305},
  [\href{https://arxiv.org/abs/1804.10378}{{\ttfamily 1804.10378}}].

\bibitem{Higaki:2013qka}
T.~Higaki, K.~Nakayama and F.~Takahashi, \emph{{Cosmological constraints on
  axionic dark radiation from axion-photon conversion in the early Universe}},
  \href{https://doi.org/10.1088/1475-7516/2013/09/030}{\emph{JCAP} {\bfseries
  09} (2013) 030}, [\href{https://arxiv.org/abs/1306.6518}{{\ttfamily
  1306.6518}}].

\bibitem{Jaeckel:2021ert}
J.~Jaeckel and W.~Yin, \emph{{Shining ALP dark radiation}},
  \href{https://doi.org/10.1103/PhysRevD.105.115003}{\emph{Phys. Rev. D}
  {\bfseries 105} (2022) 115003},
  [\href{https://arxiv.org/abs/2110.03692}{{\ttfamily 2110.03692}}].

\bibitem{Jaeckel:2021gah}
J.~Jaeckel and W.~Yin, \emph{{Using the spectrum of dark radiation as a probe
  of reheating}},
  \href{https://doi.org/10.1103/PhysRevD.103.115019}{\emph{Phys. Rev. D}
  {\bfseries 103} (2021) 115019},
  [\href{https://arxiv.org/abs/2102.00006}{{\ttfamily 2102.00006}}].

\bibitem{Schiavone:2021imu}
F.~Schiavone, D.~Montanino, A.~Mirizzi and F.~Capozzi, \emph{{Axion-like
  particles from primordial black holes shining through the Universe}},
  \href{https://doi.org/10.1088/1475-7516/2021/08/063}{\emph{JCAP} {\bfseries
  08} (2021) 063}, [\href{https://arxiv.org/abs/2107.03420}{{\ttfamily
  2107.03420}}].

\bibitem{Cui:2017ytb}
Y.~Cui, M.~Pospelov and J.~Pradler, \emph{{Signatures of Dark Radiation in
  Neutrino and Dark Matter Detectors}},
  \href{https://doi.org/10.1103/PhysRevD.97.103004}{\emph{Phys. Rev. D}
  {\bfseries 97} (2018) 103004},
  [\href{https://arxiv.org/abs/1711.04531}{{\ttfamily 1711.04531}}].

\bibitem{Gu:2021lni}
Y.~Gu, L.~Wu and B.~Zhu, \emph{{Axion dark radiation: Hubble tension and the
  Hyper-Kamiokande neutrino experiment}},
  \href{https://doi.org/10.1103/PhysRevD.105.095008}{\emph{Phys. Rev. D}
  {\bfseries 105} (2022) 095008},
  [\href{https://arxiv.org/abs/2105.07232}{{\ttfamily 2105.07232}}].

\bibitem{Braun:2015B3}
R.~Braun, T.~L. Bourke, J.~A. Green, E.~Keane and J.~Wagg, \emph{{Advancing
  Astrophysics with the Square Kilometre Array}},
  \href{https://doi.org/10.22323/1.215.0174}{\emph{PoS} {\bfseries AASKA14}
  (2015) 174}.

\bibitem{SKA}
R.~Braun et~al.
  \url{https://www.skao.int/sites/default/files/documents/d2-SKA-TEL-SKO-0000818-01_SKA1_Science_Perform.pdf},
  2017.

\bibitem{Kelley:2017vaa}
K.~Kelley and P.~J. Quinn, \emph{{A Radio Astronomy Search for Cold Dark Matter
  Axions}}, \href{https://doi.org/10.3847/2041-8213/aa808d}{\emph{Astrophys. J.
  Lett.} {\bfseries 845} (2017) L4},
  [\href{https://arxiv.org/abs/1708.01399}{{\ttfamily 1708.01399}}].

\bibitem{Sigl:2017sew}
G.~Sigl, \emph{{Astrophysical Haloscopes}},
  \href{https://doi.org/10.1103/PhysRevD.96.103014}{\emph{Phys. Rev. D}
  {\bfseries 96} (2017) 103014},
  [\href{https://arxiv.org/abs/1708.08908}{{\ttfamily 1708.08908}}].

\bibitem{Caputo:2018ljp}
A.~Caputo, C.~P.~n. Garay and S.~J. Witte, \emph{{Looking for Axion Dark Matter
  in Dwarf Spheroidals}},
  \href{https://doi.org/10.1103/PhysRevD.98.083024}{\emph{Phys. Rev. D}
  {\bfseries 98} (2018) 083024},
  [\href{https://arxiv.org/abs/1805.08780}{{\ttfamily 1805.08780}}].

\bibitem{Battye:2019aco}
R.~A. Battye, B.~Garbrecht, J.~I. McDonald, F.~Pace and S.~Srinivasan,
  \emph{{Dark matter axion detection in the radio/mm-waveband}},
  \href{https://doi.org/10.1103/PhysRevD.102.023504}{\emph{Phys. Rev. D}
  {\bfseries 102} (2020) 023504},
  [\href{https://arxiv.org/abs/1910.11907}{{\ttfamily 1910.11907}}].

\bibitem{Wang:2021hfb}
J.-W. Wang, X.-J. Bi and P.-F. Yin, \emph{{Detecting axion dark matter through
  the radio signal from Omega Centauri}},
  \href{https://doi.org/10.1103/PhysRevD.104.103015}{\emph{Phys. Rev. D}
  {\bfseries 104} (2021) 103015},
  [\href{https://arxiv.org/abs/2109.00877}{{\ttfamily 2109.00877}}].

\bibitem{Ayad:2020fzc}
A.~Ayad and G.~Beck, \emph{{Axion-like particle searches with MeerKAT and
  SKA}}, \href{https://doi.org/10.1088/1475-7516/2022/03/005}{\emph{JCAP}
  {\bfseries 03} (2022) 005},
  [\href{https://arxiv.org/abs/2010.05773}{{\ttfamily 2010.05773}}].

\bibitem{CAST:2017uph}
{\scshape CAST} collaboration, V.~Anastassopoulos et~al., \emph{{New CAST Limit
  on the Axion-Photon Interaction}},
  \href{https://doi.org/10.1038/nphys4109}{\emph{Nature Phys.} {\bfseries 13}
  (2017) 584--590}, [\href{https://arxiv.org/abs/1705.02290}{{\ttfamily
  1705.02290}}].

\bibitem{Payez:2014xsa}
A.~Payez, C.~Evoli, T.~Fischer, M.~Giannotti, A.~Mirizzi and A.~Ringwald,
  \emph{{Revisiting the SN1987A gamma-ray limit on ultralight axion-like
  particles}}, \href{https://doi.org/10.1088/1475-7516/2015/02/006}{\emph{JCAP}
  {\bfseries 02} (2015) 006},
  [\href{https://arxiv.org/abs/1410.3747}{{\ttfamily 1410.3747}}].

\bibitem{Mirizzi:2009nq}
A.~Mirizzi, J.~Redondo and G.~Sigl, \emph{{Constraining resonant photon-axion
  conversions in the Early Universe}},
  \href{https://doi.org/10.1088/1475-7516/2009/08/001}{\emph{JCAP} {\bfseries
  08} (2009) 001}, [\href{https://arxiv.org/abs/0905.4865}{{\ttfamily
  0905.4865}}].

\bibitem{Cicoli:2014bfa}
M.~Cicoli, J.~P. Conlon, M.~C.~D. Marsh and M.~Rummel, \emph{{3.55 keV photon
  line and its morphology from a 3.55 keV axionlike particle line}},
  \href{https://doi.org/10.1103/PhysRevD.90.023540}{\emph{Phys. Rev. D}
  {\bfseries 90} (2014) 023540},
  [\href{https://arxiv.org/abs/1403.2370}{{\ttfamily 1403.2370}}].

\bibitem{Geha:2008zr}
M.~Geha, B.~Willman, J.~D. Simon, L.~E. Strigari, E.~N. Kirby, D.~R. Law
  et~al., \emph{{The Least Luminous Galaxy: Spectroscopy of the Milky Way
  Satellite Segue 1}},
  \href{https://doi.org/10.1088/0004-637X/692/2/1464}{\emph{Astrophys. J.}
  {\bfseries 692} (2009) 1464--1475},
  [\href{https://arxiv.org/abs/0809.2781}{{\ttfamily 0809.2781}}].

\bibitem{Simon_2011}
J.~D. Simon, M.~Geha, Q.~E. Minor, G.~D. Martinez, E.~N. Kirby, J.~S. Bullock
  et~al., \emph{A {COMPLETE} {SPECTROSCOPIC} {SURVEY} {OF} {THE} {MILKY} {WAY}
  {SATELLITE} {SEGUE} 1: {THE} {DARKEST} {GALAXY}},
  \href{https://doi.org/10.1088/0004-637x/733/1/46}{\emph{The Astrophysical
  Journal} {\bfseries 733} (may, 2011) 46}.

\bibitem{Calmet:2020rpx}
X.~Calmet and F.~Kuipers, \emph{{Bounds on very weakly interacting ultra light
  scalar and pseudoscalar dark matter from quantum gravity}},
  \href{https://doi.org/10.1140/epjc/s10052-020-8350-7}{\emph{Eur. Phys. J. C}
  {\bfseries 80} (2020) 781},
  [\href{https://arxiv.org/abs/2008.06243}{{\ttfamily 2008.06243}}].

\bibitem{DES:2020mpv}
{\scshape DES} collaboration, A.~Chen et~al., \emph{{Constraints on dark matter
  to dark radiation conversion in the late universe with DES-Y1 and external
  data}}, \href{https://doi.org/10.1103/PhysRevD.103.123528}{\emph{Phys. Rev.
  D} {\bfseries 103} (2021) 123528},
  [\href{https://arxiv.org/abs/2011.04606}{{\ttfamily 2011.04606}}].

\bibitem{Bringmann:2018jpr}
T.~Bringmann, F.~Kahlhoefer, K.~Schmidt-Hoberg and P.~Walia, \emph{{Converting
  nonrelativistic dark matter to radiation}},
  \href{https://doi.org/10.1103/PhysRevD.98.023543}{\emph{Phys. Rev. D}
  {\bfseries 98} (2018) 023543},
  [\href{https://arxiv.org/abs/1803.03644}{{\ttfamily 1803.03644}}].

\bibitem{Geringer-Sameth:2014yza}
A.~Geringer-Sameth, S.~M. Koushiappas and M.~Walker, \emph{{Dwarf galaxy
  annihilation and decay emission profiles for dark matter experiments}},
  \href{https://doi.org/10.1088/0004-637X/801/2/74}{\emph{Astrophys. J.}
  {\bfseries 801} (2015) 74},
  [\href{https://arxiv.org/abs/1408.0002}{{\ttfamily 1408.0002}}].

\bibitem{Caputo:2018vmy}
A.~Caputo, M.~Regis, M.~Taoso and S.~J. Witte, \emph{{Detecting the Stimulated
  Decay of Axions at RadioFrequencies}},
  \href{https://doi.org/10.1088/1475-7516/2019/03/027}{\emph{JCAP} {\bfseries
  03} (2019) 027}, [\href{https://arxiv.org/abs/1811.08436}{{\ttfamily
  1811.08436}}].

\bibitem{Alonso-Alvarez:2019ssa}
G.~Alonso-\'Alvarez, R.~S. Gupta, J.~Jaeckel and M.~Spannowsky, \emph{{On the
  Wondrous Stability of ALP Dark Matter}},
  \href{https://doi.org/10.1088/1475-7516/2020/03/052}{\emph{JCAP} {\bfseries
  03} (2020) 052}, [\href{https://arxiv.org/abs/1911.07885}{{\ttfamily
  1911.07885}}].

\bibitem{Raffelt:1987im}
G.~Raffelt and L.~Stodolsky, \emph{{Mixing of the Photon with Low Mass
  Particles}}, \href{https://doi.org/10.1103/PhysRevD.37.1237}{\emph{Phys. Rev.
  D} {\bfseries 37} (1988) 1237}.

\bibitem{Grossman:2002by}
Y.~Grossman, S.~Roy and J.~Zupan, \emph{{Effects of initial axion production
  and photon axion oscillation on type Ia supernova dimming}},
  \href{https://doi.org/10.1016/S0370-2693(02)02448-6}{\emph{Phys. Lett. B}
  {\bfseries 543} (2002) 23--28},
  [\href{https://arxiv.org/abs/hep-ph/0204216}{{\ttfamily hep-ph/0204216}}].

\bibitem{Carenza:2021alz}
P.~Carenza, C.~Evoli, M.~Giannotti, A.~Mirizzi and D.~Montanino,
  \emph{{Turbulent axion-photon conversions in the Milky~Way}},
  \href{https://doi.org/10.1103/PhysRevD.104.023003}{\emph{Phys. Rev. D}
  {\bfseries 104} (2021) 023003},
  [\href{https://arxiv.org/abs/2104.13935}{{\ttfamily 2104.13935}}].

\bibitem{Colafrancesco:2006he}
S.~Colafrancesco, S.~Profumo and P.~Ullio, \emph{{Detecting dark matter WIMPs
  in the Draco dwarf: A multi-wavelength perspective}},
  \href{https://doi.org/10.1103/PhysRevD.75.023513}{\emph{Phys. Rev. D}
  {\bfseries 75} (2007) 023513},
  [\href{https://arxiv.org/abs/astro-ph/0607073}{{\ttfamily
  astro-ph/0607073}}].

\bibitem{McDaniel:2017ppt}
A.~McDaniel, T.~Jeltema, S.~Profumo and E.~Storm, \emph{{Multiwavelength
  Analysis of Dark Matter Annihilation and RX-DMFIT}},
  \href{https://doi.org/10.1088/1475-7516/2017/09/027}{\emph{JCAP} {\bfseries
  09} (2017) 027}, [\href{https://arxiv.org/abs/1705.09384}{{\ttfamily
  1705.09384}}].

\bibitem{Natarajan:2015hma}
A.~Natarajan, J.~E. Aguirre, K.~Spekkens and B.~S. Mason, \emph{{Green Bank
  Telescope Constraints on Dark Matter Annihilation in Segue I}},
  \href{https://arxiv.org/abs/1507.03589}{{\ttfamily 1507.03589}}.

\bibitem{Arshakian:2008cx}
T.~G. Arshakian, R.~Beck, M.~Krause and D.~Sokoloff, \emph{{Evolution of
  magnetic fields in galaxies and future observational tests with the Square
  Kilometre Array}},
  \href{https://doi.org/10.1051/0004-6361:200810964}{\emph{Astron. Astrophys.}
  {\bfseries 494} (2009) 21},
  [\href{https://arxiv.org/abs/0810.3114}{{\ttfamily 0810.3114}}].

\bibitem{Regis:2014koa}
M.~Regis, L.~Richter, S.~Colafrancesco, S.~Profumo, W.~J.~G. de~Blok and
  M.~Massardi, \emph{{Local Group dSph radio survey with ATCA \textendash{} II.
  Non-thermal diffuse emission}},
  \href{https://doi.org/10.1093/mnras/stv127}{\emph{Mon. Not. Roy. Astron.
  Soc.} {\bfseries 448} (2015) 3747--3765},
  [\href{https://arxiv.org/abs/1407.5482}{{\ttfamily 1407.5482}}].

\bibitem{2008IAUS..248..164T}
A.~R. {Taylor}, \emph{{The Square Kilometre Array}},  in \emph{A Giant Step:
  from Milli- to Micro-arcsecond Astrometry} (W.~J. {Jin}, I.~{Platais} and
  M.~A.~C. {Perryman}, eds.), vol.~248, pp.~164--169, July, 2008,
  \href{https://doi.org/10.1017/S1743921308018954}{DOI}.

\bibitem{2012PhDT.......123B}
J.~D. {Bregman}, \emph{{System Design and Wide-field Imaging Aspects of
  Synthesis Arrays with Phased Array Stations}}, Ph.D. thesis, University of
  Groningen, Netherlands, Dec., 2012.

\bibitem{2021ExA....51....1B}
J.~G. {Bij de Vaate}, D.~I.~L. {de Villiers}, D.~B. {Davidson} and W.~A. {van
  Cappellen}, \emph{{Expanding the field of view: station design for the AAMID
  SKA radio telescope}},
  \href{https://doi.org/10.1007/s10686-020-09682-9}{\emph{Experimental
  Astronomy} {\bfseries 51} (Feb., 2021) 1--16}.

\bibitem{Chen:2021rea}
Z.~Chen, Y.-L.~S. Tsai and Q.~Yuan, \emph{{Sensitivity of SKA to dark matter
  induced radio emission}},
  \href{https://doi.org/10.1088/1475-7516/2021/09/025}{\emph{JCAP} {\bfseries
  09} (2021) 025}, [\href{https://arxiv.org/abs/2105.00776}{{\ttfamily
  2105.00776}}].

\bibitem{VLA}
\url{https://science.nrao.edu/facilities/vla/docs/manuals/obsguide/modes/mosaicking}.

\bibitem{1988A&A...202..316C}
T.~J. {Cornwell}, \emph{{Radio-interferometric imaging of very large
  objects.}}, {\emph{\aap} {\bfseries 202} (Aug., 1988) 316--321}.

\bibitem{Ghara:2015mab}
R.~Ghara, T.~R. Choudhury and K.~K. Datta, \emph{{21-cm signature of the first
  sources in the Universe: Prospects of detection with SKA}},
  \href{https://doi.org/10.1093/mnras/stw953}{\emph{Mon. Not. Roy. Astron.
  Soc.} {\bfseries 460} (2016) 827--843},
  [\href{https://arxiv.org/abs/1511.07448}{{\ttfamily 1511.07448}}].

\bibitem{Kar:2019cqo}
A.~Kar, S.~Mitra, B.~Mukhopadhyaya and T.~R. Choudhury, \emph{{Heavy dark
  matter particle annihilation in dwarf spheroidal galaxies: radio signals at
  the SKA telescope}},
  \href{https://doi.org/10.1103/PhysRevD.101.023015}{\emph{Phys. Rev. D}
  {\bfseries 101} (2020) 023015},
  [\href{https://arxiv.org/abs/1905.11426}{{\ttfamily 1905.11426}}].

\bibitem{Planck:2015zrl}
{\scshape Planck} collaboration, P.~A.~R. Ade et~al., \emph{{Planck 2015
  results. XIX. Constraints on primordial magnetic fields}},
  \href{https://doi.org/10.1051/0004-6361/201525821}{\emph{Astron. Astrophys.}
  {\bfseries 594} (2016) A19},
  [\href{https://arxiv.org/abs/1502.01594}{{\ttfamily 1502.01594}}].

\bibitem{Beck_2011}
R.~Beck, \emph{Magnetic fields in galaxies},  in \emph{Space Sciences Series of
  {ISSI}}, pp.~215--230.
\newblock Springer New York, 2011.
\newblock \href{https://doi.org/10.1007/978-1-4614-5728-2_8}{DOI}.

\bibitem{Kachelriess:2021rzc}
M.~Kachelriess and J.~Tjemsland, \emph{{On the origin and the detection of
  characteristic axion wiggles in photon spectra}},
  \href{https://doi.org/10.1088/1475-7516/2022/01/025}{\emph{JCAP} {\bfseries
  01} (2022) 025}, [\href{https://arxiv.org/abs/2111.08303}{{\ttfamily
  2111.08303}}].

\bibitem{Conlon:2014xsa}
J.~P. Conlon and F.~V. Day, \emph{{3.55 keV photon lines from axion to photon
  conversion in the Milky Way and M31}},
  \href{https://doi.org/10.1088/1475-7516/2014/11/033}{\emph{JCAP} {\bfseries
  11} (2014) 033}, [\href{https://arxiv.org/abs/1404.7741}{{\ttfamily
  1404.7741}}].

\end{thebibliography}\endgroup

\end{document}